\documentclass[english,12pt, draftclsnofoot, onecolumn]{IEEEtran}
\usepackage[LGR,T1]{fontenc}
\usepackage[latin9]{inputenc}
\usepackage{color}
\usepackage{float}
\usepackage{units}
\usepackage{mathrsfs}
\usepackage{mathtools}
\usepackage{amsmath}
\usepackage{amsthm}
\usepackage{amssymb}
\usepackage{graphicx}
\usepackage{setspace}
\makeatletter

\DeclareRobustCommand{\greektext}{%
  \fontencoding{LGR}\selectfont\def\encodingdefault{LGR}}
\DeclareRobustCommand{\textgreek}[1]{\leavevmode{\greektext #1}}
\ProvideTextCommand{\~}{LGR}[1]{\char126#1}

\providecommand{\tabularnewline}{\\}
\floatstyle{ruled}
\newfloat{algorithm}{tbp}{loa}
\providecommand{\algorithmname}{Algorithm}
\floatname{algorithm}{\protect\algorithmname}

\theoremstyle{plain}
\newtheorem{thm}{\protect\theoremname}
\theoremstyle{plain}
\newtheorem{lem}[thm]{\protect\lemmaname}

\usepackage{eqnarray}
\usepackage{mathrsfs}
\usepackage{epsfig}
\usepackage[english]{babel}
\usepackage{subfigure}
\usepackage{epstopdf}
\usepackage{import}
\usepackage{color}
\usepackage{colortbl}\usepackage{cite}
\usepackage{algorithm}

\usepackage{bbm}
\usepackage{cases}
\usepackage{array}
\usepackage[english]{babel}
\usepackage{times}
\usepackage{color}
\usepackage{amsfonts}
\usepackage{psfrag}
\usepackage{fancyhdr}
 \usepackage{algorithmic}
\allowdisplaybreaks

\usepackage{babel}
\providecommand{\lemmaname}{Lemma}
\providecommand{\theoremname}{Theorem}
\usepackage{babel}
\providecommand{\lemmaname}{Lemma}
\providecommand{\theoremname}{Theorem}

\makeatother

\usepackage{babel}
\providecommand{\lemmaname}{Lemma}
\providecommand{\theoremname}{Theorem}

\begin{document}
\title{Optimized Age of Information Tail for Ultra-Reliable Low-Latency Communications
in Vehicular Networks\thanks{This work was supported in part by the Academy of Finland project
CARMA, and 6Genesis Flagship (grant no. 318927), in part by the INFOTECH
project NOOR, in part by the Office of Naval Research (ONR) under
MURI Grant N00014-19-1-2621, in part by the National Science Foundation
under Grant IIS-1633363, and in part by the Kvantum Institute strategic
project SAFARI. A preliminary conference version of this work appears
in the proceedings of IEEE GLOBECOM 2018 \cite{Abde1812:Ultra}.}}
\author{\IEEEauthorblockN{Mohamed~K.~Abdel-Aziz,~\IEEEmembership{Student Member,~IEEE},
Sumudu~Samarakoon,~\IEEEmembership{Member,~IEEE}, Chen-Feng~Liu,~\IEEEmembership{Student Member,~IEEE},
Mehdi~Bennis,~\IEEEmembership{Senior Member,~IEEE}, and Walid~Saad,~\IEEEmembership{Fellow Member,~IEEE} }\thanks{M. K. Abdel-Aziz, S. Samarakoon, C.-F. Liu, and M. Bennis are with
the Centre for Wireless Communications, University of Oulu, 90014
Oulu, Finland (e-mails: mohamed.abdelaziz@oulu.fi; sumudu.samarakoon@oulu.fi;
chen-feng.liu@oulu.fi; mehdi.bennis@oulu.fi).}\thanks{W.~Saad is with the Department of Electrical and Computer Engineering,
Virginia Polytechnic Institute and State University, Blacksburg, VA
24061, USA (e-mail: walids@vt.edu).}}
\maketitle
\begin{abstract}
While the notion of age of information (AoI) has recently been proposed
for analyzing ultra-reliable low-latency communications (URLLC), most
of the existing works have focused on the average AoI measure. Designing
a wireless network based on average AoI will fail to characterize
the performance of URLLC systems, as it cannot account for extreme
AoI events, occurring with very low probabilities. In contrast, this
paper goes beyond the average AoI to improve URLLC in a vehicular
communication network by characterizing and controlling the AoI tail
distribution. In particular, the transmission power minimization problem
is studied under stringent URLLC constraints in terms of probabilistic
AoI for both deterministic and Markovian traffic arrivals. Accordingly,
an efficient novel mapping between AoI and queue-related distributions
is proposed. Subsequently, extreme value theory (EVT) and Lyapunov
optimization techniques are adopted to formulate and solve the problem
considering both long and short packets transmissions. Simulation
results show over a two-fold improvement, in shortening the AoI distribution
tail, versus a baseline that models the maximum queue length distribution,
in addition to a tradeoff between arrival rate and AoI.
\end{abstract}


\section{Introduction\label{sec:Introduction}}

Vehicle-to-vehicle (V2V) communication will play an important role
in next-generation (5G) mobile networks and is envisioned as one of
the most promising enabler for intelligent transportation systems
\cite{Araniti2013,6G,Zeng2018}. Typically, V2V safety applications
(e.g., forward collision warning, blind spot/lane change warning,
and adaptive cruise control) are known to be \emph{time-critical},
as they rely on acquiring real-time status updates from individual
vehicles. In this regard, the European telecommunications standards
institute (ETSI) has standardized two types of safety messages: cooperative
awareness messages (CAMs) and decentralized environmental notification
messages (DENMs) \cite{etsi2011intelligent}. One key challenge for
delivering such critical and status update messages in V2V networks
is how to provide ultra-reliable and low-latency vehicular communication
links.

Indeed, achieving ultra-reliable low-latency communication represents
one of the major challenges faced by 5G and vehicular networks \cite{Mehdi_riskTail}.
In particular, a system design based on conventional average values
(e.g., latency, rate, and queue length) is not adequate to capture
the URLLC requirements, since averages often ignore the occurrence
of extreme events (e.g., high latency events) that negatively impacts
the overall performance. To overcome this challenge, one can resort
to the robust framework of \emph{extreme value theory} (EVT) which
characterizes the probability distributions of extreme events, defined
as the tail of the latency distribution or queue length \cite{EVT}.
Remarkably, the majority of the existing V2V literature focuses primarily
on average performance metrics \cite{Ashraf2016,Ikram2017,Liu2017,Mei2018},
which is not sufficient in a URLLC setting. Only a handful of recent
works have considered extreme values for vehicular networks \cite{Mouradian2016,ChenBaseline2018}.
In particular, the work in \cite{Mouradian2016} focuses on studying
large delays in vehicular networks using EVT, via simulations using
realistic mobility traces, without considering any analytical formulation.
In \cite{ChenBaseline2018}, the authors study the problem of transmit
power minimization subject to a new reliability measure in terms of
maximal queue length among all vehicle pairs. Therein, EVT was utilized
to characterize the distribution of maximal queue length over the
network.

Since V2V safety applications are time-critical, the freshness of
a vehicle's status updates is of high importance along with the low-latency
requirement \cite{champatiAoI,Champati2019statistical,AoI_First}.
A relevant metric in quantifying this freshness is the notion of \emph{age
of information} (AoI) proposed in \cite{AoI_First}. AoI is defined
as the time elapsed since the generation instant of the latest received
status update at a destination. Optimizing the AoI is fundamentally
different from delay or throughput optimization. In \cite{AoI_First},
the authors derive the minimum AoI at an optimal operating point that
lies between the extremes of maximum throughput and minimum delay.
Thus, providing quality-of-service (QoS) in terms of AoI is essential
for any time-critical application and has attracted lots of research
interest recently in various fields\footnote{An interested reader may refer to \cite{KostaAoISurvey} and \cite{sun2019ongoing}
for a comprehensive survey and references on AoI.} such as energy harvesting \cite{EnergyHarvesting1,EnergyHarvesting2},
wireless networked control systems \cite{Champati2019statistical},
and vehicular networks \cite{AoI_First,AoIBaiocchi2017,AoIAbd-Elmagid2018}.
However, except for \cite{Champati2019statistical}, these works focus
on optimizing average AoI metrics. While interesting, a system design
based on average AoI cannot enable the unique requirements of URLLC.
Instead, the AoI distribution needs to be considered especially when
dealing with time-critical V2V safety applications. Only a handful
of works investigated the distribution of AoI, e.g.,\cite{champatiAoI,Champati2019statistical,AoIDistributionInoue2018},
and \cite{PeakAoIDevassy2018}. In \cite{champatiAoI,Champati2019statistical}
the authors argue that computing an exact expression for the AoI distribution
may not always be feasible. Therefore, they opt for computing a bound
on the tail of the AoI distribution and use that bound to formulate
a tractable \textgreek{a}-relaxed upper bound minimization problem
(\textgreek{a}-UBMP) to find an optimal sampling (i.e. arrival) rate
that minimizes the AoI violation probability for a given age limit.
In \cite{AoIDistributionInoue2018}, the AoI distribution is obtained
in terms of the Laplace-Stieltjes transform (LST) and is expressed
in terms of the stationary distributions of the system delay and the
peak AoI\footnote{Therefore, considering the AoI violation in this work is a different
measure compared to the peak-AoI violation \cite{AoIDistributionInoue2018}}, where the peak AoI, proposed in \cite[Def. 3]{Costa2016}, is another
freshness metric used within the literature \cite{Costa2016,PeakAoIHuang2015,PeakAoIHe2016,PeakAoIDevassy2018}.
Finally, in \cite{PeakAoIDevassy2018}, the peak AoI violation probability
is first characterized by deriving the probability generating function
(PGF) of the peak age. Then the violation probability is obtained
through a saddlepoint approximation. However, while interesting, these
works do not consider controlling the tail of AoI violation distribution.

\subsection{Contributions}

The main contribution of this paper is a novel framework that allow
the control of the tail of AoI distribution in V2V communication networks,
thus going beyond the conventional average-based AoI. In particular,
our key goal is to enable vehicular user equipment (VUE) pairs to
minimize their transmit power while ensuring stringent latency and
reliability constraints based on a probabilistic AoI violation measure.
To capture both periodic and stochastic arrivals, we consider two
queuing systems, namely a D/G/1 and a M/G/1 queuing system. To this
end, since the AoI metric is a receiver-side metric while the transmit
power allocation occurs at the transmitter, we first derive a novel
relationship between the probabilistic AoI and the queue length of
each VUE for the D/G/1 system, and between the probabilistic AoI and
the arrival rate of each VUE, for the M/G/1 system. Moreover, in order
to constrain the exceedance over the imposed threshold, we use the
fundamental concepts of EVT to characterize the tail, and the excess
value of the vehicles' queues and arrival rates, which are then incorporated
as the statistical constraints within our transmission power minimization
problem. Furthermore, it is assumed that a roadside unit (RSU) is
used to cluster VUEs into disjoint groups in terms of their geographic
locations, thus mitigating interference and reducing the signaling
overhead between VUEs and the RSU. Since the objective function and
constraints are represented in time-averaged and steady-state forms,
Lyapunov stochastic optimization techniques \cite{LyapunovNeely2010}
are leveraged by each VUE pair to locally optimize its transmission
power subject to probabilistic AoI constraints. Knowing the instantaneous
state realization, the Lyapunov optimization framework allows each
VUE to optimize its transmission power in each time slot based on
the observed random events without considering the queue length transition
between the current and successive time slots. Moreover, the Lyapunov
framework allows optimization of long-term performance metric while
achieving network stability.

We further note that in V2V communication, the size of safety messages
are typically small (e.g., \cite{Hassanzadeh2009} and \cite{Wang2017}).
In addition, due to the high mobility of V2V networks, the small time
slot duration restricts the blocklength in each transmission. The
finite blocklength transmission will hinder the vehicles to achieve
the Shannon rate with an infinitesimal decoding error probability
\cite{Durisi2016}. In this regard, incorporating the blocklength
and decoding error probability, the authors in \cite{Polyanskiy2010}
have provided an approximated rate formula and shown a performance
gap between the Shannon rate-based design and finite blocklength regime.
As a result, the impact of short packets on V2V network performance
should also be investigated.

Therefore, in addition to the system analysis based on the Shannon
rate, we also study the power minimization problem considering the
short packet transmission which, however, yields a non-convex optimization
problem. To deal with this non-convexity, the convex-concave procedure
(CCP) is used to solve the power minimization problem. Finally, simulation
results corroborate the usefulness of EVT in characterizing the distribution
of AoI. The results also show that the proposed approach yields over
two-fold performance gains in AoI compared to two baselines: the first
baseline's scheme is concerned about the high-order statistics of
the network-wide maximal queue length but oblivious of the AoI \cite{ChenBaseline2018},
while the second baseline does not take into account the URLLC design
for taming the AoI tail. Our results also reveal an interesting tradeoff
between the arrival rate of the status updates, and the average and
worst AoI achieved by the network. Finally, the results show the existence
of a blocklength at which the violation probability of AoI is minimized.

In a nutshell, the main contributions of this work are as follows.
\begin{itemize} \item We derive and propose a novel relationship
between the probabilistic AoI and the queue dynamics of each VUE for
the D/G/1 system. (Lemma 1) \item We derive and propose a novel relationship
between the probabilistic AoI and the queue dynamics of each VUE for
the M/G/1 system. (Lemma 2,3) \item Utilizing those relationships,
we propose a novel framework to minimize VUEs' transmit power while
ensuring the probabilistic AoI constraint in terms of the derived
queue dynamics relation. \item We invoke EVT to characterize the
violation probability and Lyapunov optimization to solve the problem.
\item In the finite blocklength regime, the convex-concave procedure
(CCP) is utilized to convexify the power minimization problem. \end{itemize}

The rest of this paper is organized as follows. In Section \ref{sec:System-model},
the system model is described. The reliability constraints and the
studied problems, for both deterministic and Markovian arrival cases,
are presented in Section \ref{sec:Problem-formulation}, followed
by the proposed AoI-aware resource allocation policy in Section \ref{sec:AoI-aware-optimal-control}.
In Section \ref{sec:Numerical-results}, numerical results are presented
while conclusions are drawn in Section \ref{sec:Conclusion}.

\section{System Model\label{sec:System-model}}

\begin{figure}[t]
\centering \includegraphics[width=0.8\columnwidth]{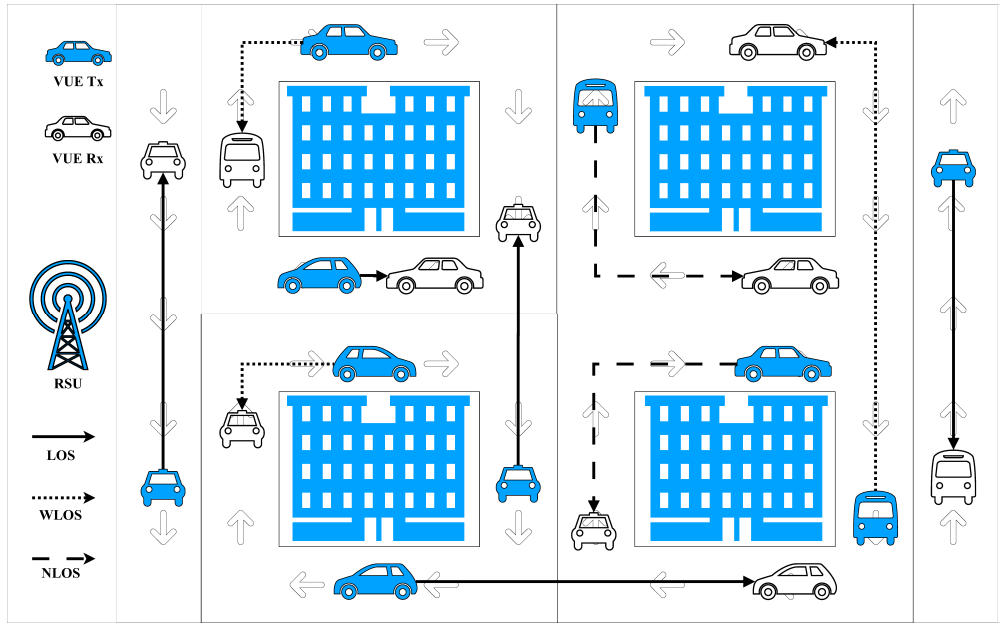}
\caption{System and path loss models of the considered V2V network.}
\label{fig:Mobility_and_Path_model}
\end{figure}

As shown in Fig.\,\ref{fig:Mobility_and_Path_model}, we consider
a V2V communication network based on a Manhattan mobility model \cite{Manhattan},
composed of a set $\mathcal{K}$ of $K$ VUE transmitter-receiver
pairs\footnote{A fixed number of vehicles represents the vehicular mobility behavior
under the assumption of equal arrival and departure flow rates.} under the coverage of a single RSU. During the entire communication
lifetime, the association of each transmitter-receiver is assumed
to be fixed. One potential application of this setup is to avoid the
rear-end collision between vehicles, where the transmitter (e.g.,
vehicle in the front) sends a collision-warning message to the receiver
(e.g., vehicle in the back) in a unicast manner\footnote{Investigating multicast schemes within the context of V2V communication
is important. However, the unicast scheme study is a challenging problem
on its own. The multicast extension is worth a dedicated study, which
is left as a future work.} \cite{ETSI_1,ETSI_2}.

We consider a slotted communication timeline indexed by $t$, and
the duration of each slot is denoted by $\tau$. Additionally, all
VUE pairs share a set $\mathcal{N}$ of $N$ orthogonal resource blocks
(RBs) with bandwidth $\omega$ per RB. We further denote the RB allocation
as $\eta_{k}^{n}(t)\in\left\{ 0,1\right\} ,\forall k\in\mathcal{K},n\in\mathcal{N}$,
where $\eta_{k}^{n}(t)=1$ indicates that RB $n$ is used by VUE pair
$k$ in time slot $t$ and $\eta_{k}^{n}(t)=0$ otherwise. The transmitter
of pair $k$ allocates a transmit power $P_{k}^{n}(t)\geq0$ over
RB $n$ to serve its receiver subject to a power constraint $\sum_{n\in\mathcal{N}}\eta_{k}^{n}(t)P_{k}^{n}(t)\leq P_{\textrm{max}}$,
where $P_{\textrm{max}}$ is the total power budget per each VUE pair.

Let $h_{kk'}^{n}(t)$ be the instantaneous channel gain, including
path loss and channel fading, from the transmitter of pair $k$ to
the receiver of pair $k'$ over RB $n$ in slot $t$. We consider
the $5.9\text{ GHz}$ carrier frequency and adopt the path loss model
in \cite{Path_loss_model}. For the path loss model, we have the line-of-sight
(LOS), weak-line-of-sight (WLOS), and non-line-of-sight (NLOS) cases.
Let us first denote an arbitrary transmitter's and an arbitrary receiver's
Euclidean coordinates as $\boldsymbol{x}=(x_{1},x_{2})\in\mathbb{R}^{2}$
and $\boldsymbol{y}=(y_{1},y_{2})\in\mathbb{R}^{2}$, respectively.
When the transmitter and receiver are on the same lane, we have an
LOS path loss value as $l_{0}\left\Vert \boldsymbol{x}-\boldsymbol{y}\right\Vert ^{-\alpha}$,
where $\left\Vert .\right\Vert $ is the $l_{2}$-norm, $l_{0}$ is
the path loss coefficient, and $\alpha$ is the path loss exponent.
Additionally, when the transmitter and receiver are located separately
on perpendicular lanes, we have the WLOS or NLOS case, depending on
the transmitter and receiver's locations. If, at least, one is near
the intersection within a distance $\mathscr{D}$, we consider the
WLOS path loss, i.e., $l_{0}\left\Vert \boldsymbol{x}-\boldsymbol{y}\right\Vert _{1}^{-\alpha}$
with the $l_{1}$-norm $\left\Vert .\right\Vert _{1}$. Otherwise,
the NLOS case, with the path loss value $l_{0}^{'}\left(|x_{1}-y_{1}|.|x_{2}-y_{2}|\right)^{-\alpha}$
and the path loss coefficient $l_{0}^{'}<l_{0}\left(\frac{\mathscr{D}}{2}\right)^{\alpha}$,
is considered. Fig.\,\ref{fig:Mobility_and_Path_model} illustrates
these three path loss cases. The Shannon data rate of VUE pair $k$
in time slot $t$ (in the unit of packets per slot) is given by\footnote{Note that \eqref{eq:Rate} is an approximation that holds when $\tau\to\infty$.}
\begin{equation}
R_{k}(t)=\frac{\omega\tau}{Z}{\textstyle \sum\limits _{n\in\mathcal{N}}}\log_{2}\left(1+\frac{P_{k}^{n}(t)h_{kk}^{n}(t)}{N_{0}\omega+I_{k}^{n}(t)}\right),\label{eq:Rate}
\end{equation}
where $Z$ is the total packet size in bits which includes the payload
of the packet as well as any headers/preambles, and $N_{0}$ is the
power spectral density of the additive white Gaussian noise. Here,
$I_{k}^{n}(t)=\sum_{k'\in\mathcal{K}/k}\eta_{k'}^{n}(t)P_{k'}^{n}(t)h_{k'k}^{n}(t)$
is the aggregate interference at the receiver of VUE pair $k$ over
RB $n$ received from other VUE pairs operating over the same RB.
We rely on a bit-pipe abstraction of the physical layer where bits
are delivered reliably at a rate equal to \eqref{eq:Rate}.

Furthermore, each VUE transmitter has a queue buffer to store the
data to be delivered to the desired receiver following a first-come
first-serve (FCFS) policy. Denoting the VUE pair $k$'s queue length
at the beginning of slot $t$ as $Q_{k}(t)$, the queue dynamics will
be given by 
\begin{equation}
Q_{k}(t+1)=\max\left(Q_{k}(t)-R_{k}(t),0\right)+A_{k}(t),\label{eq:Physical_queue}
\end{equation}
where $A_{k}(t)$ is the instantaneous packet arrival for VUE pair
$k$ during slot $t$. In order to ensure queue stability, the following
constraint needs to be satisfied
\begin{equation}
\lim_{C\to\infty}\frac{1}{C}{\textstyle \sum\limits _{t=0}^{C-1}}R_{k}(t)>\lambda\text{,}~\forall k\in\mathcal{K},\label{eq:queueStability_const}
\end{equation}
where $\lambda=\lim_{C\to\infty}\frac{1}{C}{\textstyle \sum\limits _{t=0}^{C-1}}A_{k}(t)$
is the average packet arrival rate per slot. Within this work, two
arrival processes will be considered: 1) The deterministic arrival
process, i.e., D/G/1, which account for the periodic nature of CAMs;
2) The Poisson arrival process, i.e., M/G/1, which accounts for the
packet generation triggered by random events such as a hazard on the
road\footnote{Considering last-come first serve (LCFS) or queuing model based on
M(or D)/G/1/1 and M(or D)/G/1/2{*} could provide better AoI performance
than FCFS. Different queuing models and policies could be investigated
as a future extension.}. In the next section, the AoI-based reliability constrains are formulated
for both cases.

\section{Enabling URLLC Based on Age of Information\label{sec:Problem-formulation}}

Providing real-time status updates for mission critical applications
is a key use case in V2V networks. These applications rely on the
``freshness'' of the data, which can be quantified by the concept
of AoI \cite{AoI_First} 
\begin{equation}
\Delta_{k}(T)\triangleq T-\max_{i}\left(T_{k}^{\text{A}}\left(i\right)\mid T_{k}^{\text{D}}\left(i\right)\leq T\right).\label{eq:AoICalculation}
\end{equation}
Here, $\Delta_{k}(T)$ is the AoI of VUE pair $k$ at a time instant
$T$. $T_{k}^{\text{A}}\left(i\right)$ and $T_{k}^{\text{D}}\left(i\right)$
represent the arrival and departure instants of packet $i$ of VUE
pair $k$, respectively. As a reliability requirement, we impose a
probabilistic constraint on the AoI for each VUE pair $k\in\mathcal{K}$,
as follows: 
\begin{equation}
\lim_{T\rightarrow\infty}\mathrm{Pr}\left\{ \Delta_{k}(T)>d\right\} \leq\epsilon_{k}\text{, }\forall k\in\mathcal{K},\label{eq:AoI_constraint}
\end{equation}
where $d$ is the age threshold, and $\epsilon_{k}\ll1$ is the tolerable
AoI violation probability. Consider an average arrival rate $\nicefrac{\lambda}{\tau}$
packets per second. The support of the steady state AoI distribution
(and hence the existence of the limit in \eqref{eq:AoI_constraint})
is $\left[\nicefrac{\tau}{\lambda},\infty\right)$ since AoI cannot
be less than $\frac{\tau}{\lambda}$ with this given arrival rate
\cite{champatiAoI}. To this end, since AoI is a receiver-side metric
while the transmit power allocation occurs at the transmitter, a novel
mapping between the AoI and the transmitter's queue dynamics is proposed
for both D/G/1 and M/G/1 cases.

\subsection{D/G/1 Queuing System (Deterministic Arrival)}

In D/G/1 systems, arrivals are deterministic and periodic, as in the
case of periodic CAMs in various V2V applications. Hence, the packet
arrival rate per slot, $A_{k}(t)$, is constant and denoted by $A$,
and the packet $i$'s arrival time instant will be $T_{k}^{\text{A}}\left(i\right)=\frac{i}{A}\tau$.
Note that the indices of the packets that arrive during slot $t$
satisfy $\ensuremath{i\in[tA,(t+1)A-1]}$, while the packets that
are served during the same slot will satisfy the following condition:
\begin{equation}
tA-Q_{k}(t)\leq i\leq tA-1-\max\left(Q_{k}(t)-R_{k}(t),0\right).\label{eq:served_indices}
\end{equation}

In \cite{champatiAoI}, it is shown that for a given age limit $d_{\text{D}}$
with $\frac{A}{\tau}\geq\frac{1}{d_{\text{D}}}$\footnote{This condition is to ensure that the process is not undersampled.
If $A/\tau$ is less than \emph{$1/d_{\text{D}}$,} the arrival rate
would be too low to maintain the target age threshold $d_{\text{D}}$
\cite{champatiAoI}.}, the steady state distribution of AoI for a D/G/1 queue can be characterized
as 
\begin{equation}
\lim_{T\rightarrow\infty}\mathrm{Pr}\Bigl\{\Delta_{k}(T)>d_{\text{D}}\Bigr\}=\lim_{T\rightarrow\infty}\mathrm{Pr}T_{k}^{\text{D}}({\color{blue}\hat{\imath}})>T,\label{eq:ZubaidyTheorem}
\end{equation}
where $\hat{\imath}\triangleq\lceil\frac{A}{\tau}(T-d_{\text{D}})\rceil$
is the index of the packet that first arrives at or after time $T-d_{\text{D}}$,
and $d_{\text{D}}$ is the age threshold for the D/G/1 system. Next,
in Lemma \ref{lem:Assuming-that-packet }, we derive a mapping between
the steady state distribution of the departure instant of a given
packet and the queue length.
\begin{lem}
\label{lem:Assuming-that-packet }Given that $T$ is observed at the
beginning of each slot $t+1$, i.e. $T=\tau(t+1)$, then 
\[
\mathrm{Pr}\left\{ T_{k}^{\text{D}}(\hat{\imath})>\tau(t+1)\right\} \leq\mathrm{Pr}\left\{ Q_{k}(t)>R_{k}(t)-\psi\right\} ,
\]
where $\psi=2-(\frac{d_{\text{D}}}{\tau}-1)A$.
\end{lem}
\begin{IEEEproof}
See Appendix \ref{sec:AppendixA}.
\end{IEEEproof}
Combining \eqref{eq:ZubaidyTheorem} and Lemma \ref{lem:Assuming-that-packet },
a sufficient condition for the probabilistic constraint \eqref{eq:AoI_constraint}
to hold in the slotted system can be written as 
\begin{equation}
\lim_{C\to\infty}\frac{1}{C}{\textstyle \sum\limits _{t=0}^{C-1}}\mathrm{Pr}\left\{ Q_{k}(t)>R_{k}(t)-\psi\right\} \leq\epsilon_{k}\text{, }\forall k\in\mathcal{K},\label{eq:equiv_queue_constraint}
\end{equation}
where the time average on $t$ represents the steady-state in the
slotted system.

As previously mentioned, enabling URLLC requires the characterization
of the tail of the AoI distribution. Therefore, we utilize the fundamental
concepts of EVT \cite{EVT} to investigate the event $Q_{k}(t)>R_{k}(t)-\psi$.
The \emph{Pickands\textendash Balkema\textendash de Haan theorem}
for threshold violation of a random variable \cite[Theorem 4.1]{EVT}
states that if a random variable $Q$ has a cumulative distribution
function (CDF) denoted by $F_{Q}(q)$, and has a threshold value $\delta$.
Then, as the threshold $\delta$ closely approaches $F_{Q}^{-1}(1)$,
the conditional CDF of the excess value $X=Q-\delta>0$ denoted as
$F_{X\text{|}Q>\delta}(x)=\mathrm{Pr}\left\{ (Q-\delta)<x\rvert Q>\delta\right\} $,
can be approximated by 
\[
G\left(x;\sigma,\xi\right)=\begin{cases}
1-(\max\{1+\frac{\xi x}{\sigma},0\})^{-\frac{1}{\xi}}, & \text{if }\xi\neq0,\\
1-e^{-\frac{x}{\sigma}}, & \xi=0.
\end{cases}
\]
Here, $G\left(x;\sigma,\xi\right)$ is the generalized Pareto distribution
(GPD) whose mean and variance are $\frac{\sigma}{1-\xi}$ and $\frac{\sigma^{2}}{(1-\xi)^{2}(1-2\xi)}$,
respectively. Note that the value of the scale parameter $\sigma$
is threshold-dependent, except in the case when the shape parameter
$\xi=0$ \cite{EVT}. Also, note that a very low threshold $\delta$
is likely to violate the asymptotic basis of the model, leading to
a bias; a very high threshold $\delta$ will generate few excesses
with which the model can be estimated, leading to high variance \cite{EVT}.
Moreover, the characteristics of the GPD depend on the scale parameter
$\sigma>0$ and the shape parameter $\xi<\frac{1}{2}$.

The \emph{Pickands\textendash Balkema\textendash de Haan theorem}
states that, for a sufficiently high threshold $\delta$, the distribution
function of the excess value can be approximated by a GPD. In this
regard, for \eqref{eq:equiv_queue_constraint}, we define the conditional
excess queue value of each VUE pair $k\in\mathcal{K}$ at time slot
$t$ as $X_{k}^{\text{D}}(t)|_{Q_{k}(t)>R_{k}(t)-\psi}=Q_{k}(t)-R_{k}(t)+\psi$.
Thus, we can approximate the mean and the variance of $X_{k}^{\text{D}}(t)$
as 
\begin{align}
\mathbb{E}\left[X_{k}^{\text{D}}(t)|Q_{k}(t)>R_{k}(t)-\psi\right] & \approx\frac{\sigma_{k}}{1-\xi_{k}},\label{eq:Excess_expectation}\\
\text{Var}\left[X_{k}^{\text{D}}(t)|Q_{k}(t)>R_{k}(t)-\psi\right] & \approx\frac{\sigma_{k}^{2}}{(1-\xi_{k})^{2}(1-2\xi_{k})},\label{eq:Excess_variance}
\end{align}
with a scale parameter $\sigma_{k}$ and a shape parameter $\xi_{k}$.
Note that the smaller the $\sigma_{k}$ and $\xi_{k}$, the smaller
the mean value and variance of the GPD. Hence, in order to constraint
the exceedance over the imposed threshold, we further impose thresholds
on the scale and the shape parameters, i.e., $\sigma_{k}\leq\sigma_{k}^{\textrm{th}}$
and $\xi_{k}\leq\xi_{k}^{\textrm{th}}$\footnote{The thresholds on the scale and shape parameters are application dependent
and they reflect how much excess can be allowed within the system
\cite{ChenEVT}.} \cite{Chen-fengMEC,ChenEVT}. Subsequently, applying both parameter
thresholds and $\text{Var}\left(X_{k}^{\text{D}}\right)=\mathbb{E}\left[(X_{k}^{\text{D}})^{2}\right]-\mathbb{E}\left[X_{k}^{\text{D}}\right]^{2}$
to \eqref{eq:Excess_expectation} and \eqref{eq:Excess_variance},
we impose constraints for the time-averaged mean and second moment
of the conditional excess queue value, i.e., 
\begin{align}
\hspace{-1em}\bar{X}_{k}^{\text{D}} & =\lim_{C\rightarrow\infty}\frac{1}{C}{\textstyle \sum\limits _{t=0}^{C-1}}\mathbb{E}\left[X_{k}^{\text{D}}(t)\rvert Q_{k}(t)>R_{k}(t)-\psi\right]\leq H,\label{eq:GPDMean}\\
\hspace{-1em}\bar{Y}_{k}^{\text{D}} & =\lim_{C\rightarrow\infty}\frac{1}{C}{\textstyle \sum\limits _{t=0}^{C-1}}\mathbb{E}\left[Y_{k}^{\text{D}}(t)\rvert Q_{k}(t)>R_{k}(t)-\psi\right]\leq B,\label{eq:GPD2moment}
\end{align}
where $H=\frac{\sigma_{k}^{\textrm{th}}}{1-\xi_{k}^{\textrm{th}}}$,
$B=\frac{2(\sigma_{k}^{\textrm{th}})^{2}}{(1-\xi_{k}^{\textrm{th}})(1-2\xi_{k}^{\textrm{th}})}$
and $Y_{k}^{\text{D}}(t)\coloneqq\left[X_{k}^{\text{D}}(t)\right]^{2}$.

By denoting the RB and power allocation vectors as $\boldsymbol{\eta}(t)=\left[\eta_{k}^{n}(t)\right]_{k\in\mathcal{K}}^{n\in\mathcal{N}}$
and $\boldsymbol{P}(t)=\left[P_{k}^{n}(t)\right]_{k\in\mathcal{K}}^{n\in\mathcal{N}},\forall t$,
respectively, the network-wide transmit power minimization problem
is formulated as follows: \begin{subequations}\label{First_optimization_problem}
\begin{align}
\mathbb{P}_{\text{D}}:\min_{\boldsymbol{\eta}(t),\boldsymbol{P}(t)} & ~~{\textstyle \sum\limits _{k\in\mathcal{K}}}{\textstyle \sum\limits _{n\in\mathcal{N}}}\bar{P}_{k}^{n}\nonumber \\
\text{s.t.} & ~~\eqref{eq:queueStability_const},\eqref{eq:equiv_queue_constraint},\eqref{eq:GPDMean}\text{, and }\eqref{eq:GPD2moment},\nonumber \\
 & ~~{\textstyle \sum\limits _{n\in\mathcal{N}}}\eta_{k}^{n}(t)P_{k}^{n}(t)\leq P_{\textrm{max}},~\forall k\in\mathcal{K},\label{eq:PowerConst_1}\\
 & ~~0\leq P_{k}^{n}(t)\leq P_{\textrm{max}},\text{ }\forall t,\,k\in\mathcal{K},\,n\in\mathcal{N},\label{eq:PowerConst_2}\\
 & ~~\eta_{k}^{n}(t)\in\{0,1\},~\forall t,\,k\in\mathcal{K},\,n\in\mathcal{N},\label{eq:RBConst}
\end{align}
\end{subequations}where $\bar{P}_{k}^{n}=\lim\limits _{C\to\infty}\frac{1}{C}\sum_{t=0}^{C-1}P_{k}^{n}(t)$
is the time-averaged transmit power of VUE pair $k$ over RB $n$.
In the following subsection, a similar problem is formulated and analyzed
for the scenario considering the M/G/1 queuing system.

\subsection{M/G/1 Queuing System (Stochastic Arrival)}

In M/G/1 systems, arrivals are Markovian (Poisson), which captures
scenarios in which the packet generation in vehicles is triggered
by a random event such as a road hazard, or sudden change in speed.
Therefore, the probability of $I$ packet arrivals within the slot
duration $\tau$ is 
\begin{equation}
\mathrm{Pr}\left\{ I\text{ packet arrivals within the slot duration }\tau\right\} =e^{-\lambda}\frac{\lambda^{I}}{I!},\label{eq:Poisson}
\end{equation}
where $\lambda$ is the average packet arrival rate per slot. In the
following Lemma, we present a key insight regarding the steady state
distribution of AoI for the M/G/1 queue, following similar procedures
used in \cite{champatiAoI} for the D/G/1 case.
\begin{lem}
\label{lem:Given-an-M/G/1}For an M/G/1 queuing system with an age
threshold $d_{\text{M}}$ and $T<\infty$, 
\[
\mathrm{Pr}\left\{ \Delta_{k}(T)>d_{\text{M}}\right\} =e^{-\frac{\lambda d_{\text{M}}}{\tau}}+\mathrm{Pr}T_{k}^{\text{D}}({\color{blue}\hat{\imath}})>T\left(1-e^{-\frac{\lambda d_{\text{M}}}{\tau}}\right),
\]
where $\hat{\imath}$ is the index of the packet that first arrives
at or after time $T-d_{\text{M}}$, and $d_{\text{M}}$ is the age
threshold for the M/G/1 system.
\end{lem}
\begin{IEEEproof}
See Appendix \ref{sec:AppendixB}.
\end{IEEEproof}
Next, in Lemma \ref{lem:MapMG1} we propose a mapping between the
steady state distribution of $T_{k}^{\text{D}}(\hat{\imath})$ and
the arrival rate $A_{k}(t)$.
\begin{lem}
\label{lem:MapMG1} Given that $T$ is observed at the beginning of
each slot $t$, i.e. $T=\tau t$, then 
\[
\mathrm{Pr}\left\{ T_{k}^{\text{D}}(\hat{\imath})>T\right\} =\mathrm{Pr}\left\{ A_{k}(t)>R_{k}(t)\right\} .
\]
\end{lem}
\begin{IEEEproof}
See Appendix \ref{sec:AppendixC}.
\end{IEEEproof}
Combining the results of Lemma \ref{lem:Given-an-M/G/1} and \ref{lem:MapMG1},
the probabilistic constraint \eqref{eq:AoI_constraint} can be rewritten
as, 
\begin{equation}
\lim_{C\to\infty}\frac{1}{C}{\textstyle \sum\limits _{t=0}^{C-1}}\mathrm{Pr}\left\{ A_{k}(t)>R_{k}(t)\right\} \leq E_{k}\text{, }\forall k\in\mathcal{K},\label{eq:equiv_Constraint_MG1}
\end{equation}
where $E_{k}=\frac{\epsilon_{k}-e^{-\frac{\lambda d_{\text{M}}}{\tau}}}{1-e^{-\frac{\lambda d_{\text{M}}}{\tau}}}$,
and $d_{\text{M}}\geq-\frac{\tau\ln\epsilon_{k}}{\lambda}$ to ensure
$E_{k}\geq0$. Similar to the previous subsection, we investigate
the event $A_{k}(t)>R_{k}(t)$ and study the tail behavior of AoI
using the \emph{Pickands\textendash Balkema\textendash de Haan theorem}
for threshold violation \cite{EVT}. This yields two constraints for
the time-averaged mean and second moment of the conditional excess
value, as follows: 
\begin{align}
\hspace{-1em}\bar{X}_{k}^{\text{M}} & =\lim_{C\rightarrow\infty}\frac{1}{C}{\textstyle \sum\limits _{t=0}^{C-1}}\mathbb{E}\left[X_{k}^{\text{M}}(t)\rvert A_{k}(t)>R_{k}(t)\right]\leq H,\label{eq:GPDMean-MG1}\\
\hspace{-1em}\bar{Y}_{k}^{\text{M}} & =\lim_{C\rightarrow\infty}\frac{1}{C}{\textstyle \sum\limits _{t=0}^{C-1}}\mathbb{E}\left[Y_{k}^{\text{M}}(t)\rvert A_{k}(t)>R_{k}(t)\right]\leq B,\label{eq:GPD2moment-MG1}
\end{align}
where $X_{k}^{\text{M}}(t)|_{A_{k}(t)>R_{k}(t)}=A_{k}(t)-R_{k}(t)$,
and $Y_{k}^{\text{M}}(t)\coloneqq\left[X_{k}^{\text{M}}(t)\right]^{2}$.
In this regard, the network wide transmit power minimization problem
for the M/G/1 case is given by \begin{subequations}\label{First_optimization_problem-MG1}
\begin{align*}
\mathbb{P}_{\text{M}}:\min_{\boldsymbol{\eta}(t),\boldsymbol{P}(t)} & ~~{\textstyle \sum\limits _{k\in\mathcal{K}}}{\textstyle \sum\limits _{n\in\mathcal{N}}}\bar{P}_{k}^{n}\\
\text{s.t.} & ~~\eqref{eq:queueStability_const},\eqref{eq:PowerConst_1},\eqref{eq:PowerConst_2},\eqref{eq:RBConst},\eqref{eq:equiv_Constraint_MG1},\eqref{eq:GPDMean-MG1},\text{ and }\eqref{eq:GPD2moment-MG1}.
\end{align*}
\end{subequations}

\section{AoI-Aware Resource Allocation Using Lyapunov Optimization\label{sec:AoI-aware-optimal-control}}

Since the objective function and the constrains are represented in
a time-average and steady-state forms, and in order to find the optimal
resource $\boldsymbol{\eta}(t)$ and power $\boldsymbol{P}(t)$ allocation
vectors of both deterministic and Markovian arrivals corresponding
to problems $\mathbb{P}_{\text{D}}$ and $\mathbb{P}_{\text{M}}$,
we invoke techniques from Lyapunov stochastic optimization \cite{LyapunovNeely2010}.
Later on, we will study the impact of short packets and their effect
on the problem formulation.

\subsection{Deterministic Arrivals}

\label{Sec: Deterministic Arrivals}

To solve $\mathbb{P}_{\text{D}}$, we first rewrite the probabilistic
constraint in \eqref{eq:equiv_queue_constraint} as a time-averaged
constraint, so it could be utilized by the Lyapunov stochastic optimization,
as follows: 
\begin{equation}
\lim_{C\rightarrow\infty}\frac{1}{C}{\textstyle \sum\limits _{t=0}^{C-1}}R_{k}(t)\mathbbm{1}\left\{ Q_{k}(t)>R_{k}(t)-\psi\right\} \leq\bar{\epsilon}_{k},\label{eq:modified_queueConstraint}
\end{equation}
where $\bar{\epsilon}_{k}=\lim_{C\rightarrow\infty}\frac{1}{C}\sum_{t=0}^{C-1}R_{k}(t)\epsilon_{k}$
is the product of the time-averaged rate and the tolerance value $\epsilon_{k}$,
and $\mathbbm{1}\left\{ .\right\} $ is the indicator function. Using
Lyapunov optimization, the time-averaged constraints \eqref{eq:GPDMean},
\eqref{eq:GPD2moment}, \eqref{eq:queueStability_const}, and \eqref{eq:modified_queueConstraint}
can be satisfied by converting them into virtual queues and maintain
their stability \cite{LyapunovNeely2010}, i.e. the lower bound of
these constraints are considered as the arrival rate to the virtual
queue, while the upper bounds are considered as its service rate.
In this regard, we introduce the corresponding virtual queues with
the dynamics shown in \eqref{eq:Virtual1}\textendash \eqref{eq:Virtual4}.
\begin{figure*}
\begin{flalign}
J_{k}^{(X)}(t+1) & =\max\left(J_{k}^{(X)}(t)+(X_{k}^{\text{D}}(t)-H)\mathbbm{1}\left\{ Q_{k}(t)>R_{k}(t)-\psi\right\} ,0\right),\label{eq:Virtual1}\\
J_{k}^{(Y)}(t+1) & =\max\left(J_{k}^{(Y)}(t)+(Y_{k}^{\text{D}}(t)-B)\mathbbm{1}\left\{ Q_{k}(t)>R_{k}(t)-\psi\right\} ,0\right),\label{eq:Virtual2}\\
J_{k}^{(R)}(t+1) & =\max\left(J_{k}^{(R)}(t)-R_{k}(t)+A,0\right),\label{eq:Virtual3}\\
J_{k}^{(Q)}(t+1) & =\max\left(J_{k}^{(Q)}(t)+R_{k}(t)\mathbbm{1}\left\{ Q_{k}(t)>R_{k}(t)-\psi\right\} -R_{k}(t)\epsilon_{k},0\right).\label{eq:Virtual4}
\end{flalign}
\noindent\makebox[1\linewidth]{%
\rule{0.84\paperwidth}{0.4pt}%
} 
\end{figure*}

For notation simplicity, let $\boldsymbol{J}(t)=\bigl[J_{k}^{(X)}(t),J_{k}^{(Y)}(t),J_{k}^{(Q)}(t),J_{k}^{(R)}(t),Q_{k}(t):k\in\mathcal{K}\bigr]$
denotes the combined physical and virtual queues vector. Then, in
order to maintain the stability of $\boldsymbol{J}(t)$, we use the
conditional Lyapunov drift-plus-penalty for time slot $t$, which
can be expressed as 
\begin{equation}
\mathbb{E}\biggl[\mathscr{L}\left(\boldsymbol{J}(t+1)\right)-\mathscr{L}\left(\boldsymbol{J}(t)\right)+\sum_{k\in\mathcal{K}}\sum_{n\in\mathcal{N}}VP_{k}^{n}(t)\rvert\boldsymbol{J}(t)\biggr],\label{eq:Lyapunov_drift}
\end{equation}
where $\mathscr{L}\left(\boldsymbol{J}(t)\right)=\frac{\boldsymbol{J^{\text{T}}}(t)\boldsymbol{J}(t)}{2}$
is the Lyapunov function with $\boldsymbol{J^{\text{T}}}(t)$ being
the transpose of $\boldsymbol{J}(t)$, and $V\geq0$ is a parameter
that controls the tradeoff between optimal transmit power and queue
stability. By calculating the Lyapunov drift and leveraging the fact
that $\left(\max\left(a-b,0\right)+c\right)^{2}\leq a^{2}+b^{2}+c^{2}-2a(b-c)\text{, }\forall a,b,c\geq0$,
and $\left(\max\left(x,0\right)\right)^{2}\leq x^{2}$, on \eqref{eq:Physical_queue}
and \eqref{eq:Virtual1}\textendash \eqref{eq:Virtual4}, an upper
bound on \eqref{eq:Lyapunov_drift} can be obtained as 
\begin{flalign}
\eqref{eq:Lyapunov_drift} & \leq\mathfrak{C}^{\text{D}}+\mathbb{E}\Biggl[\sum_{k\in\mathcal{K}}\biggl(\Bigl(J_{k}^{(Q)}(t)-J_{k}^{(X)}(t)-2\left(Q_{k}(t)+\psi\right)^{3}\nonumber \\
 & -\left(2J_{k}^{(Y)}(t)+1\right)\left(Q_{k}(t)+\psi\right)\Bigr)\cdot\mathbbm{1}\left\{ Q_{k}(t)>R_{k}(t)-\psi\right\} \nonumber \\
 & -\left(J_{k}^{(R)}(t)+A+Q_{k}(t)+J_{k}^{(Q)}(t)\cdot\epsilon_{k}\right)\biggr)R_{k}(t)\nonumber \\
 & +\sum_{k\in\mathcal{K}}\sum_{n\in\mathcal{N}}VP_{k}^{n}(t)\rvert\boldsymbol{J}(t)\Biggr].\label{eq:Lyapunov_bound}
\end{flalign}
Here, $\mathfrak{C}^{\text{D}}=A^{2}+\left(J_{k}^{(R)}(t)+Q_{k}(t)\right)A+\left(1+\frac{1}{2}\epsilon_{k}^{2}\right)R_{k}^{2}(t)+\biggl(\frac{1}{2}\left(\left(Q_{k}(t)+\psi\right)^{4}+H^{2}+B^{2}\right)+\left(Q_{k}(t)+\psi\right)^{2}\left(\frac{1}{2}-B+J_{k}^{(Y)}(t)\right)+\left(Q_{k}(t)+\psi\right)\left(J_{k}^{(X)}(t)-H\right)-J_{k}^{(X)}(t)H-J_{k}^{(Y)}(t)B+\Bigl(H+2B\left(Q_{k}(t)+\psi\right)\Bigr)R_{k}(t)+\Bigl(1+3\left(Q_{k}(t)+\psi\right)^{2}-B+J_{k}^{(Y)}(t)-\epsilon_{k}\Bigr)R_{k}^{2}(t)-2\left(Q_{k}(t)+\psi\right)R_{k}^{3}(t)+\frac{1}{2}R_{k}^{4}(t)\biggr)\cdot\mathbbm{1}\left\{ Q_{k}(t)>R_{k}(t)-\psi\right\} $
is a bounded term that does not affect the system performance. Note
that the solution to problem $\mathbb{P}_{\text{D}}$ can be obtained
by minimizing the upper bound in \eqref{eq:Lyapunov_bound} in each
slot $t$ \cite{LyapunovNeely2010}, i.e., 
\begin{align*}
\hat{\mathbb{P}}_{\text{D}}:\min_{\boldsymbol{\eta}(t),\mathbf{P}(t)} & {\textstyle \sum\limits _{k\in\mathcal{K}}}\Biggl[{\textstyle \sum\limits _{n\in\mathcal{N}}}VP_{k}^{n}(t)-\biggl[J_{k}^{(R)}(t)+A+Q_{k}(t)\\
 & +J_{k}^{(Q)}(t)\cdot\epsilon_{k}+\Bigl(-J_{k}^{(Q)}(t)+J_{k}^{(X)}(t)\\
 & +(2J_{k}^{(Y)}(t)+1)(Q_{k}(t)+\psi)+2(Q_{k}(t)+\psi)^{3}\Bigr)\\
 & \cdot\mathbbm{1}\left\{ Q_{k}(t)>R_{k}(t)-\psi\right\} \biggr]R_{k}(t)\Biggr]\\
\text{s.t. } & \eqref{eq:PowerConst_1}\text{-}\eqref{eq:RBConst}.
\end{align*}

\subsection{Markovian Arrivals}

To solve $\mathbb{P}_{\text{M}}$, we first rewrite \eqref{eq:equiv_Constraint_MG1}
as 
\begin{equation}
\lim_{C\rightarrow\infty}\frac{1}{C}{\textstyle \sum\limits _{t=0}^{C-1}}R_{k}(t)\mathbbm{1}\left\{ A_{k}(t)>R_{k}(t)\right\} \leq\bar{E}_{k},\label{eq:modified_MG1_constraint}
\end{equation}
where $\bar{E}_{k}=\lim_{C\rightarrow\infty}\frac{1}{C}\sum_{t=0}^{C-1}R_{k}(t)E_{k}$.
Following the same procedures as in Section \ref{Sec: Deterministic Arrivals},
the time-averaged constraints \eqref{eq:GPDMean-MG1}, \eqref{eq:GPD2moment-MG1},
\eqref{eq:queueStability_const}, and \eqref{eq:modified_MG1_constraint}
are converted into virtual queues with the dynamics shown in \eqref{eq:Virtual1-MG1}\textendash \eqref{eq:Virtual4-MG1}.
\begin{figure*}
\begin{flalign}
M_{k}^{(X)}(t+1) & =\max\left(M_{k}^{(X)}(t)+(X_{k}^{\text{M}}(t)-H)\mathbbm{1}\left\{ A_{k}(t)>R_{k}(t)\right\} ,0\right),\label{eq:Virtual1-MG1}\\
M_{k}^{(Y)}(t+1) & =\max\left(M_{k}^{(Y)}(t)+(Y_{k}^{\text{M}}(t)-B)\mathbbm{1}\left\{ A_{k}(t)>R_{k}(t)\right\} ,0\right),\label{eq:Virtual2-MG1}\\
M_{k}^{(R)}(t+1) & =\max\left(M_{k}^{(R)}(t)-R_{k}(t)+A_{k}(t),0\right),\label{eq:Virtual3-MG1}\\
M_{k}^{(Q)}(t+1) & =\max\left(M_{k}^{(Q)}(t)+R_{k}(t)\mathbbm{1}\left\{ A_{k}(t)>R_{k}(t)\right\} -R_{k}(t)E_{k},0\right).\label{eq:Virtual4-MG1}
\end{flalign}
\noindent\makebox[1\linewidth]{%
\rule{0.84\paperwidth}{0.4pt}%
} 
\end{figure*}

Denoting $\boldsymbol{M}(t)=\bigl[M_{k}^{(X)}(t),M_{k}^{(Y)}(t),M_{k}^{(Q)}(t),M_{k}^{(R)}(t),Q_{k}(t):k\in\mathcal{K}\bigr]$
as the combined physical and virtual queue vector, the conditional
Lyapunov drift-plus-penalty for slot $t$ is given by 
\begin{equation}
\mathbb{E}\biggl[\mathscr{L}\left(\boldsymbol{M}(t+1)\right)-\mathscr{L}\left(\boldsymbol{M}(t)\right)+\sum_{k\in\mathcal{K}}\sum_{n\in\mathcal{N}}VP_{k}^{n}(t)\rvert\boldsymbol{M}(t)\biggr].\label{eq:Lyapunov_drift-MG1}
\end{equation}

By calculating the Lyapunov drift and leveraging the fact that $\left(\max\left(a-b,0\right)+c\right)^{2}\leq a^{2}+b^{2}+c^{2}-2a(b-c)\text{, }\forall a,b,c\geq0$,
and $\left(\max\left(x,0\right)\right)^{2}\leq x^{2}$, on \eqref{eq:Physical_queue}
and \eqref{eq:Virtual1-MG1}\textendash \eqref{eq:Virtual4-MG1},
an upper bound on $\eqref{eq:Lyapunov_drift-MG1}$ can be obtained
as follows: 
\begin{flalign}
\eqref{eq:Lyapunov_drift-MG1} & \leq\mathfrak{C}^{\text{M}}+\mathbb{E}\Biggl[\sum_{k\in\mathcal{K}}\biggl(\Bigl(M_{k}^{(Q)}(t)-M_{k}^{(X)}(t)-2A_{k}^{3}(t)\nonumber \\
 & -\left(2M_{k}^{(Y)}(t)+1\right)A_{k}(t)\Bigr)\cdot\mathbbm{1}\left\{ A_{k}(t)>R_{k}(t)\right\} \nonumber \\
 & -\left(M_{k}^{(R)}(t)+A_{k}(t)+Q_{k}(t)+M_{k}^{(Q)}(t)\cdot E_{k}\right)\biggr)R_{k}(t)\nonumber \\
 & +\sum_{k\in\mathcal{K}}\sum_{n\in\mathcal{N}}VP_{k}^{n}(t)\rvert\boldsymbol{M}(t)\Biggr].\label{eq:Lyapunov_bound-MG1}
\end{flalign}
Here, $\mathfrak{C}^{\text{M}}=\frac{1}{2}\left(A_{k}^{2}(t)+A_{k}(t)\right)+\Bigl(M_{k}^{(R)}(t)+Q_{k}(t)\Bigr)A_{k}(t)+\left(1+\frac{1}{2}E_{k}^{2}\right)R_{k}^{2}(t)+\biggl(\frac{1}{2}\left(A_{k}^{4}(t)+H^{2}+B^{2}\right)+A_{k}^{2}(t)\left(\frac{1}{2}-B+M_{k}^{(Y)}(t)\right)+A_{k}(t)\left(M_{k}^{(X)}(t)-H\right)-M_{k}^{(X)}(t)H-M_{k}^{(Y)}(t)B+\Bigl(H+2BA_{k}(t)\Bigr)R_{k}(t)+\Bigl(1+3A_{k}^{2}(t)-B+M_{k}^{(Y)}(t)-E_{k}\Bigr)R_{k}^{2}(t)-2A_{k}(t)R_{k}^{3}(t)+\frac{1}{2}R_{k}^{4}(t)\biggr)\cdot\mathbbm{1}\left\{ A_{k}(t)>R_{k}(t)\right\} $
is a bounded term that does not affect the system performance. The
solution to problem $\mathbb{P}_{\text{M}}$ can be obtained by minimizing
the upper bound in \eqref{eq:Lyapunov_bound-MG1} in each slot $t$
\cite{LyapunovNeely2010}, i.e., 
\begin{align*}
\hat{\mathbb{P}}_{\text{M}}:\min_{\boldsymbol{\eta}(t),\mathbf{P}(t)} & {\textstyle \sum\limits _{k\in\mathcal{K}}}\Biggl[{\textstyle \sum\limits _{n\in\mathcal{N}}}VP_{k}^{n}(t)-\biggl[M_{k}^{(R)}(t)+A_{k}(t)+Q_{k}(t)\\
 & +M_{k}^{(Q)}(t).E_{k}+\Bigl(-M_{k}^{(Q)}(t)+M_{k}^{(X)}(t)\\
 & +(2M_{k}^{(Y)}(t)+1)A_{k}(t)+2A_{k}^{3}(t)\Bigr)\\
 & \cdot\mathbbm{1}\left\{ A_{k}(t)>R_{k}(t)\right\} \biggr]R_{k}(t)\Biggr]\\
\text{s.t. } & \eqref{eq:PowerConst_1}\text{--}\eqref{eq:RBConst}.
\end{align*}

To solve $\hat{\mathbb{P}}_{\text{D}}$ and $\hat{\mathbb{P}}_{\text{M}}$
each time slot $t$, the RSU needs full global channel state information
(CSI) and queue state information (QSI). This is clearly impractical
for vehicular networks since frequently exchanging fast-varying local
information between the RSU and VUEs can yield a significant unacceptable
overhead. To alleviate the information exchange burden, we utilize
a two-timescale resource allocation mechanism which is performed in
two stages. Therein, RBs for each VUE pair are centrally allocated
over a long timescale at the RSU whereas each VUE pair minimizes its
transmit power over a short timescale. In the next subsection, a more
about the two-stage resource allocation mechanism is presented.

\subsection{Two-Stage Resource Allocation}

\subsubsection{Spectral Clustering and RB Allocation at the RSU}

It can be noted that the co-channel transmission of nearby VUE pairs
can lead to severe interference. In order to avoid the interference
from nearby VUEs, the RSU first clusters VUE pairs into $g>1$ disjoint
groups, in which the nearby VUE pairs are allocated to the same group,
and then the RSU orthogonally allocates all RBs to the VUE pairs in
each group. Vehicle clustering is done by means of spectral clustering
due to its ease of implementation and its efficient solvability by
standard linear algebra methods \cite{Spectral_Clustering}. We adopt
the VUE clustering and RB allocation technique as in \cite{ChenBaseline2018},
denoting $\boldsymbol{v}_{k}\in\mathbb{R}^{2}$ as the Euclidean coordinate
of the midpoint of VUE transmitter-receiver pairs $k$. Here, we use
a distance-based Gaussian similarity matrix $\boldsymbol{F}$ to represent
the geographic proximity information, in which the $(k,k')$-th element
is defined as 
\begin{align*}
f_{kk'}\coloneqq\begin{cases}
e^{-\left\Vert \boldsymbol{v}_{k}-\boldsymbol{v}_{k'}\right\Vert ^{2}/\gamma^{2}}, & \left\Vert \boldsymbol{v}_{k}-\boldsymbol{v}_{k'}\right\Vert \leq\phi,\\
0, & \text{otherwise},
\end{cases}
\end{align*}
where $\phi$ captures the neighborhood size, while $\gamma$ controls
the impact of the neighborhood size. Subsequently, $\boldsymbol{F}$
is used to group VUE pairs using spectral clustering as shown in Algorithm
\ref{alg:Spectral_Clustering}. The most expensive step, in terms
of computational complexity, within Algorithm \ref{alg:Spectral_Clustering}
is the computation of the eigenvalues/eigenvectors of $\boldsymbol{I}-\boldsymbol{D}^{-\frac{1}{2}}\boldsymbol{F}\boldsymbol{D}^{-\frac{1}{2}}$
(step 3) which has a complexity $\mathcal{O}\left(K^{3}\right)$.
The overall computational complexity of Algorithm \ref{alg:Spectral_Clustering}
is $\mathcal{O}\left(K^{3}\right)$ \cite{tsironis2013accurate}.
Note that, the number of VUE pairs $K$ under the coverage of a single
RSU is typically small. Hence, the computational complexity of the
proposed solution will be reasonable in practice. 
\begin{algorithm}[t]
\begin{algorithmic}[1]

\STATE \textbf{Inputs: }the Euclidean coordinate $\boldsymbol{v}_{k}$
of each VUE pair $k$, and the number of groups $g$.

\STATE  Calculate matrix $\boldsymbol{F}$ and the diagonal matrix
$\boldsymbol{D}$ with the diagonal $d_{j}=\sum_{q=1}^{K}f_{jq}$.

\STATE  Let $\boldsymbol{U}=\left[\boldsymbol{u}_{1},\cdots,\boldsymbol{u}_{g}\right]$
in which $\boldsymbol{u}_{g}$ is the eigenvector of the $g$-th smallest
eigenvalue of $\boldsymbol{I}-\boldsymbol{D}^{-\frac{1}{2}}\boldsymbol{F}\boldsymbol{D}^{-\frac{1}{2}}$.

\STATE Use $k$-means clustering approach to cluster $K$ normalized
row vectors (which represent $K$ VUE pairs) of matrix $\boldsymbol{U}$
into $g$ groups.

\STATE \textbf{Output: }$K$ VUE pairs distributed among $g$ groups.

\end{algorithmic}

\caption{\label{alg:Spectral_Clustering}Spectral Clustering for VUE Grouping}
\end{algorithm}

After forming the groups, the RSU orthogonally allocates RBs to the
VUEs inside the group. Hereafter, we denote $\mathcal{N}_{k}$ as
the set of RBs that are allocated for VUE pair $k\in\mathcal{K}$.
Moreover, to reduce the signaling overhead due to frequent information
exchange between the RSU and VUE pairs, it is assumed that VUE clustering
and RB allocation are performed in a longer time scale, i.e., every
$T_{0}\gg1$ time slots since the vehicles' geographic location do
not change significantly during the slot duration $\tau$ (i.e., coherence
time of fading channels). Therefore, VUE pairs send their locations
to the RSU only once every $T_{0}$ slots instead of every slot\footnote{Since there is no communication between VUEs and RSU during these
$T_{0}$ time slots, the performance of V2V communication observed
in a single cell represents the average V2V communication performance
over multiple cells.}.

\subsubsection{Transmit Power Allocation at the VUE}

Since VUE pair $k$ can only use the set $\mathcal{N}_{k}$ of allocated
RBs for the communication, we modify the power allocation and RB usage
constraints, i.e., \eqref{eq:PowerConst_1}\textendash \eqref{eq:RBConst},
$\forall k\in\mathcal{K}$, as 
\begin{align}
\begin{cases}
\sum\limits _{n\in\mathcal{N}_{k}}P_{k}^{n}(t)\leq P_{\textrm{max}},~\forall t,\\
P_{k}^{n}(t)\geq0,~\forall t,\,n\in\mathcal{N}_{k},\\
P_{k}^{n}(t)=0,~\forall t,\,n\notin\mathcal{N}_{k}.
\end{cases}\label{eq:PowerConst_3}
\end{align}
Note that the RBs in $\mathcal{N}_{k}$ are reused by VUE transmitters
in different groups. For tractability, we approximately treat the
aggregate interference as a constant term $I_{0}$\footnote{Since interference is from multiple distant VUEs in other clusters,
there is no dominant and significantly-dynamic interference signal.
We approximately treat the aggregate interference power as a constant
term.} and rewrite the transmission rate in \eqref{eq:Rate} as 
\begin{equation}
R_{k}(t)\approx\frac{\omega\tau}{Z}{\textstyle \sum\limits _{n\in\mathcal{N}_{k}}}\log_{2}\left(1+\frac{P_{k}^{n}(t)h_{kk}^{n}(t)}{N_{0}\omega+I_{0}}\right).\label{eq:rate_2}
\end{equation}
Subsequently, applying \eqref{eq:PowerConst_3} and \eqref{eq:rate_2}
to $\hat{\mathbb{P}}_{\text{D}}$ and $\hat{\mathbb{P}}_{\text{M}}$,
the VUE transmitter of each VUE pair $k$ locally allocates its transmit
power by solving the following convex optimization problem $\mathbb{P}_{1}$
in each slot $t$ 
\begin{align*}
\mathbb{P}_{1}:\min_{P_{k}^{n}(t)} & ~~{\textstyle \sum\limits _{n\in\mathcal{N}_{k}}}VP_{k}^{n}(t)-\Im_{k}(t)\log_{2}\left(1+\frac{P_{k}^{n}(t)h_{kk}^{n}(t)}{N_{0}\omega+I_{0}}\right)\\
\text{subject to} & ~~\eqref{eq:PowerConst_3},
\end{align*}
where for $\hat{\mathbb{P}}_{\text{D}}$ that represents the D/G/1
system, $\Im_{k}(t)=\frac{\tau\omega}{Z}\Bigl[J_{k}^{(R)}(t)+A+Q_{k}(t)+J_{k}^{(Q)}(t)\epsilon_{k}+\bigl(-J_{k}^{(Q)}(t)+J_{k}^{(X)}(t)+(2J_{k}^{(Y)}(t)+1)(Q_{k}(t)+\psi)+2(Q_{k}(t)+\psi)^{3}\bigr)\cdot\mathbbm{1}\left\{ Q_{k}(t)>R_{k}(t)-\psi\right\} \Bigr]$,
while for $\hat{\mathbb{P}}_{\text{M}}$ that represents the M/G/1
system, $\Im_{k}(t)=\frac{\tau\omega}{Z}\Bigl[M_{k}^{(R)}(t)+A_{k}(t)+Q_{k}(t)+M_{k}^{(Q)}(t)E_{k}+\bigl(-M_{k}^{(Q)}(t)+M_{k}^{(X)}(t)+(2M_{k}^{(Y)}(t)+1)A_{k}(t)+2A_{k}^{3}(t)\bigr)\cdot\mathbbm{1}\left\{ A_{k}(t)>R_{k}(t)\right\} \Bigr]$.
Based on the Karush-Kuhn-Tucker (KKT) conditions, the optimal VUE
transmit power $P_{k}^{n*}(t),\forall n\in\mathcal{N}_{k},$ of $\mathbb{P}_{1}$
satisfies 
\[
\frac{\Im_{k}(t)h_{kk}^{n}(t)}{(N_{0}\omega+I_{0}+P_{k}^{n*}(t)h_{kk}^{n}(t))\ln2}=V+\zeta,
\]
if $\frac{\Im_{k}(t)h_{kk}^{n}(t)}{(N_{0}\omega+I_{0})\ln2}>V+\zeta.$
Otherwise, $P_{k}^{n*}(t)=0.$ Moreover, the Lagrange multiplier $\zeta$
is $0$ if $\sum_{n\in\mathcal{N}_{k}}P_{k}^{n*}(t)<P_{\textrm{max}}$,
and we have $\sum_{n\in\mathcal{N}_{k}}P_{k}^{n}(t)=P_{\textrm{max}}$
when $\zeta>0$ . Note that, given a small value of $V$, the derived
power $P_{k}^{n*}(t)$ provides a sub-optimal solution to problems
$\mathbb{P}_{\text{D}}$ and $\mathbb{P}_{\text{M}}$, whose optimal
solution is asymptotically obtained by increasing $V$.

\subsection{Impact of Finite Blocklength}

Due to the high mobility feature in V2V networks, the small time slot
duration $\tau$ restricts the blocklength in each transmission. This
obstructs vehicles from achieving the Shannon rate $\eqref{eq:rate_2}$
with an infinitesimal decoding error probability \cite{Polyanskiy2010}.
In consequence, the ultra-low packet loss probability cannot be ensured.
Hence, depending on the Shannon rate $\eqref{eq:rate_2}$ may be optimistic
for designing and optimizing V2V networks. According to \cite{DBLP:journals/corr/abs-1904-10442},
finite blocklength performance can be characterized by several techniques.
one possibility is to fix the transmission rate and study the exponential
decay of the error probability as the blocklength grows. This technique
is referred to as \emph{``error exponent analysis''}. An alternative
analysis of the finite blocklength performance follows from fixing
the decoding error probability and studying the maximum transmission
rate as a function of the blocklength. This technique is referred
to as ``normal approximation'' \cite{Polyanskiy2010}. Taking into
account this practical concern in finite blocklength transmission
and following the normal approximation approach\footnote{The accuracy of the normal approximation approach within our framework
is verified in Section \ref{sec:Numerical-results}.}, the transmission rate $R_{k}(t)$ (in the unit of packet per slot)
can be reformulated as \cite{Polyanskiy2010} 
\begin{equation}
R_{k}(t)=\frac{\omega\tau}{Z}{\textstyle \sum\limits _{n\in\mathcal{N}_{k}}}\biggl(\log_{2}\left(1+\rho_{k}\right)-\sqrt{\frac{\nu_{k}}{L}}Q^{-1}(\varepsilon)+\frac{\log_{2}L}{2L}\biggr),\label{eq:ShortPacket_rate}
\end{equation}
where $\rho_{k}=\frac{P_{k}^{n}(t)h_{kk}^{n}(t)}{N_{0}\omega+I_{0}}$
is the signal-to-interference-plus-noise ratio (SINR)\footnote{Interference is treated as noise in our work. Moreover, different
decoders could be used that have a performance very close to the normal
approximation, i.e. extended Bose, Chaudhuri, and Hocquenghem (eBCH)
codes with ordered statistics decoder (OSD) \cite{FBLCodes}.} at the receiver of VUE pair $k$, $\nu_{k}=\frac{\rho_{k}(2+\rho_{k})}{\left(1+\rho_{k}\right)^{2}}\log_{2}^{2}\text{e}$
is the channel dispersion, $L$ is the blocklength, $\varepsilon$
is the desired block error probability, and $Q^{-1}(.)$ is the inverse
Gaussian $Q$ function \cite{Polyanskiy2010}. Note that, \eqref{eq:ShortPacket_rate}
implies that, in order to maintain the desired block error probability
$\varepsilon$ for a given blocklength $L$, a penalty is paid on
the Shannon rate $\eqref{eq:rate_2}$ that is proportional to $\frac{1}{\sqrt{L}}$.
Accordingly, the queue dynamics \eqref{eq:Physical_queue} is rewritten
as
\[
Q_{k}(t+1)=\max\left(Q_{k}(t)-S_{k}(t),0\right)+A_{k}(t),
\]
 where $S_{k}(t)$ is the service rate of VUE pair $k$ during slot
$t$, which is equal to \eqref{eq:ShortPacket_rate} with probability
$1-\varepsilon$, and equal to zero with probability $\varepsilon$.
Moreover, in our considered system, the blocklength $L$ is determined
by the bandwidth $\omega$ and the time slot duration $\tau$ as per
$L=\omega\tau$. Moreover, as $L\to\infty$, \eqref{eq:ShortPacket_rate}
asymptotically converges to the Shannon capacity $\eqref{eq:rate_2}$.

By replacing \eqref{eq:rate_2} with \eqref{eq:ShortPacket_rate}
in $\mathbb{P}_{1}$, $\mathbb{P}_{1}$ can be rewritten as 
\begin{align*}
\mathbb{P}_{2}:\min_{P_{k}^{n}(t)} & ~~{\textstyle \sum\limits _{n\in\mathcal{N}_{k}}}VP_{k}^{n}(t)-\Im_{k}(t)\log_{2}\left(1+\rho_{k}\right)\\
 & ~~+\frac{\Im_{k}(t)}{\sqrt{L}}(\log_{2}\text{e})Q^{-1}(\varepsilon)\sqrt{\frac{\rho_{k}(2+\rho_{k})}{\left(1+\rho_{k}\right)^{2}}}\\
\text{subject to} & ~~\eqref{eq:PowerConst_3},
\end{align*}
yielding a non-convex objective function. However, $VP_{k}^{n}(t)-\Im_{k}(t)\log_{2}\left(1+\rho_{k}\right)$
is a convex function whereas $\frac{\Im_{k}(t)}{\sqrt{L}}(\log_{2}\text{e})Q^{-1}(\varepsilon)\sqrt{\frac{\rho_{k}(2+\rho_{k})}{\left(1+\rho_{k}\right)^{2}}}$
is a concave function. Therefore, $\mathbb{P}_{2}$ belongs to the
family of difference of convex (DC) programming problems. In this
regard, the convex-concave procedure (CCP) provides an iterative and
tractable procedure which converges to a locally optimal solution
\cite{CCP}.

\subsubsection{Convex-Concave Procedure}

We first define $G\left(\boldsymbol{P}_{k}\right)=-\frac{\Im_{k}(t)}{\sqrt{L}}(\log_{2}\text{e})Q^{-1}(\varepsilon)\sqrt{\frac{\rho_{k}(2+\rho_{k})}{\left(1+\rho_{k}\right)^{2}}}$
where $\boldsymbol{P}_{k}$ denotes the transmit power allocation
vector for VUE pair $k$. Then, we select an initial feasible power
allocation $x_{0}$ and convexify $G\left(\boldsymbol{P}_{k}\right)$
using its first order Taylor approximation as follows: 
\begin{align*}
\hat{G}\left(\boldsymbol{P}_{k};x_{0}\right) & \triangleq G\left(x_{0}\right)+\nabla G\left(x_{0}\right)^{\text{T}}\left(\boldsymbol{P}_{k}-x_{0}\right),\\
\nabla G\left(x_{0}\right) & =-\frac{\Im_{k}(t)}{\sqrt{L}}(\log_{2}\text{e})Q^{-1}(\varepsilon)\cdot\biggl[\left(1+\rho_{k}\right)^{-2}\Bigl(\rho_{k}(2+\rho_{k})\Bigr)^{-\frac{1}{2}}\biggr]\biggl[\frac{h_{kk}^{n}(t)}{N_{0}\omega+I_{0}}\biggr]\Biggl|_{\boldsymbol{P}_{k}=x_{0}}.
\end{align*}
By substituting $G\left(\boldsymbol{P}_{k}\right)$ by $\hat{G}\left(\boldsymbol{P}_{k};x_{0}\right)$
in $\mathbb{P}_{2}$, $\mathbb{P}_{2}$ can be rewritten as follows:
\begin{align*}
\mathbb{P}_{3}:\min_{P_{k}^{n}(t)} & ~~{\textstyle \sum\limits _{n\in\mathcal{N}_{k}}}\left(V-\nabla G\left(x_{0}\right)\right)P_{k}^{n}(t)\\
 & ~~-\Im_{k}(t)\log_{2}\left(1+\rho_{k}\right)+\mathscr{C}\\
\text{subject to} & ~~\eqref{eq:PowerConst_3},
\end{align*}
where $\mathscr{C}=\nabla G\left(x_{0}\right)^{\text{T}}\cdot x_{0}-G\left(x_{0}\right)$
is constant for a given $x_{0}$. It is worth noting that $\mathbb{P}_{3}$
can be solved by following the same steps used in solving $\mathbb{P}_{1}$,
by simply replacing $V$ by $V-\nabla G\left(x_{0}\right)$, and it
can be solved in the same way. At each iteration, $\mathbb{P}_{3}$
is solved for a given $x_{0}$, and the optimal solution is used to
replace $x_{0}$ for the next iteration. This procedure is repeated
until a stopping criterion is satisfied. One reasonable stopping criterion
is that the improvement in the objective value is less than some threshold
$\delta$. The steps of the CCP algorithm are shown in Algorithm \ref{alg:CCP}.
Note that the computation complexity of Algorithm \ref{alg:CCP} arises
from step 4 when solving $\mathbb{P}_{3}$. $\mathbb{P}_{3}$ is a
water filling problem whose worst case complexity is $\mathcal{O}\left(N_{k}^{2}\right)$
where $N_{k}$ is the total number of RBs that are allocated for VUE
pair $k$. Therefore, the computational complexity per iteration of
Algorithm 2 is $\mathcal{O}\left(N_{k}^{2}\right)$.

\begin{algorithm}[t]
\begin{algorithmic}[1]

\STATE Initialize a feasible point $x_{j}$ to problem $\mathbb{P}_{2}$
with $j=0$.

\STATE \textbf{repeat}

\begin{ALC@g} \STATE Convexify $G\left(\boldsymbol{P}_{k}\right)$
by $\hat{G}\left(\boldsymbol{P}_{k};x_{j}\right).$

\STATE Solve $\mathbb{P}_{3}$ and denote the optimal solution as
$x^{*}(j)$.

\STATE Update $x_{j+1}=x^{*}(j)$ and $j\leftarrow j+1$.

\end{ALC@g}\STATE \textbf{until }Stopping criterion is satisfied.

\end{algorithmic}

\caption{\label{alg:CCP}CCP for solving $\mathbb{P}_{2}$}
\end{algorithm}

\section{Simulation results and analysis\label{sec:Numerical-results}}

\begin{table}[t]
\caption{Simulation parameters \cite{ChenBaseline2018}.\label{tab:sim_par}}
\centering{}%
\begin{tabular}{|c|c||c|c|}
\hline 
\textbf{Parameter}  & \textbf{Value}  & \textbf{Parameter}  & \textbf{Value}\tabularnewline
\hline 
\hline 
$N$  & $20$  & $H$  & $0.05$\tabularnewline
\hline 
$\omega$  & $180$ KHz  & $B$  & $0.0033$\tabularnewline
\hline 
$\tau$  & $3$ ms  & $g$  & $10$\tabularnewline
\hline 
$P_{\textrm{max}}$  & $23$ dBm  & $\gamma$  & $30$ m\tabularnewline
\hline 
$Z$  & $500$ Byte  & $\phi$  & $150$ m\tabularnewline
\hline 
$N_{0}$  & $-174$ dBm/Hz  & $T_{0}$  & $100$\tabularnewline
\hline 
Arrival rate  & $0.5$ Mbps  & $\alpha$  & $1.61$\tabularnewline
\hline 
$d_{\text{D}}$  & $30$ ms  & $\mathscr{D}$  & $15$ m\tabularnewline
\hline 
$d_{\text{M}}$  & $60$ ms  & $l_{0}$  & $-68.5$ dB\tabularnewline
\hline 
$\epsilon_{k}$  & $0.001$  & $l_{0}^{'}$  & $-54.5$ dB\tabularnewline
\hline 
$\psi$  & $-3.25$  & $\varepsilon$ & $10^{-5}$\tabularnewline
\hline 
$V$  & $0$  & $L$ & 550\tabularnewline
\hline 
\end{tabular}
\end{table}

In our simulations, we use a $250\times250\text{ m}^{2}$ area Manhattan
mobility model as in \cite{Ikram2017,ChenBaseline2018}. The average
vehicle speed is $60\text{ km/h}$, and the distance between the transmitter
and receiver of each VUE pair varies with time. However, the average
distance is maintained as $15\text{ m}$. Unless stated otherwise,
the remaining parameters are listed in Table \ref{tab:sim_par}. The
performance of our proposed solution is compared to two baselines,
where \emph{Baseline 1} is \cite{ChenBaseline2018}, where a power
minimization is considered subject to a reliability measure in terms
of maximal queue length, and \emph{Baseline 2} is a variant of our
model, where the EVT constraints are not considered. Results are collected
over a large number of independent runs.

\subsection{Validation of the Tail Distribution Modeling Using EVT}

\begin{figure}
\begin{centering}
\centering\includegraphics[width=0.8\columnwidth]{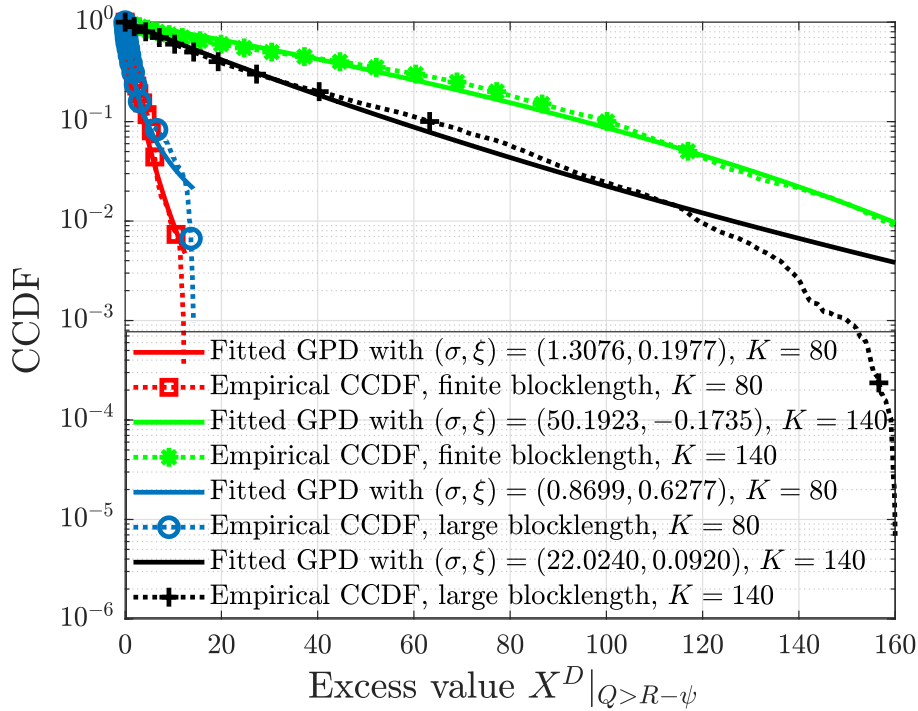}
\par\end{centering}
\caption{CCDF of the exceedance value fitted to GPD for various VUEs densities
$K$ with deterministic arrivals, finite blocklength case with $L=550$.\textcolor{blue}{\label{fig:fittingDG1}}}
\end{figure}

\begin{figure}
\begin{centering}
\centering\includegraphics[width=0.8\columnwidth]{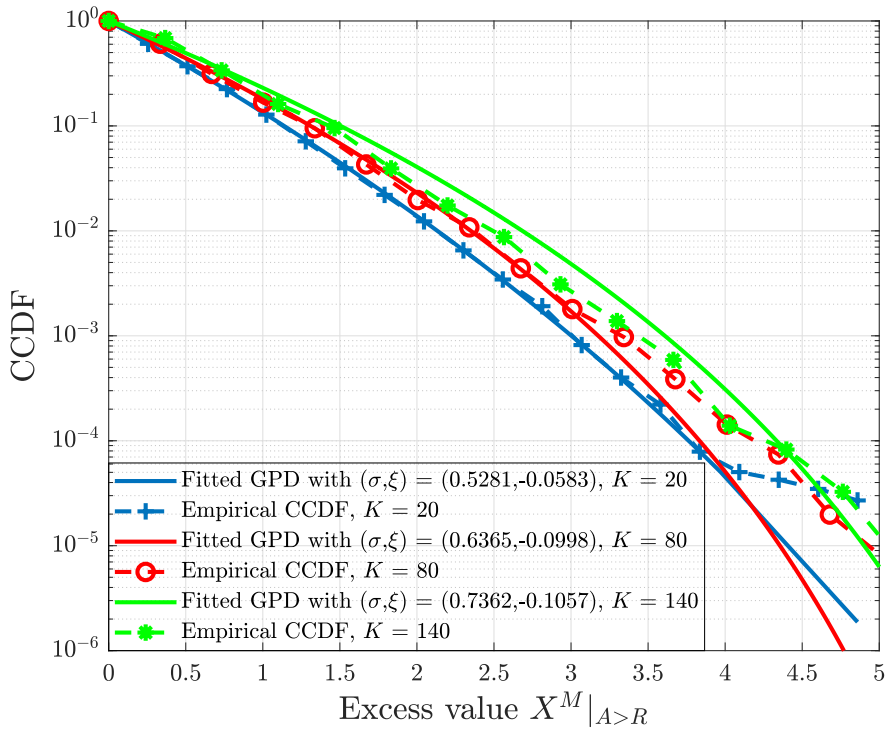}
\par\end{centering}
\caption{CCDF of the exceedance value fitted to GPD for various VUEs densities
$K$ with Markovian arrivals, large blocklength case.\textcolor{blue}{\label{fig:fittingMG1}}}
\end{figure}

In Fig.\,\ref{fig:fittingDG1} and \ref{fig:fittingMG1}, we verify
the accuracy of using EVT to characterize the distribution of the
excess value $X_{k}^{\text{D}}(t)|_{Q_{k}(t)>R_{k}(t)-\psi}=Q_{k}(t)-R_{k}(t)+\psi$
and $X_{k}^{\text{M}}(t)|_{A_{k}(t)>R_{k}(t)}=A_{k}(t)-R_{k}(t)$
for both large and finite blocklengths. Fig.\,\ref{fig:fittingDG1}
and \ref{fig:fittingMG1} show, respectively, the complementary cumulative
distribution functions (CCDF) of the generalized Pareto distribution
(GPD) and of the actual threshold violation for both deterministic
and Markovian arrival processes. An accurate fitting can be noted,
which verifies the accuracy of using EVT to characterize the distribution
of the excess value. However, due to the limitations of simulations,
the fitting becomes less accurate at higher exceedance values. For
a clear representation, the curves of the Markovian case for finite
blocklengths are omitted from Fig.\,\ref{fig:fittingMG1} as they
overlap with the curves for the large blocklength case.

\subsection{Validation of Normal Approximation}

\begin{figure}
\begin{centering}
\centering\includegraphics[width=0.8\columnwidth]{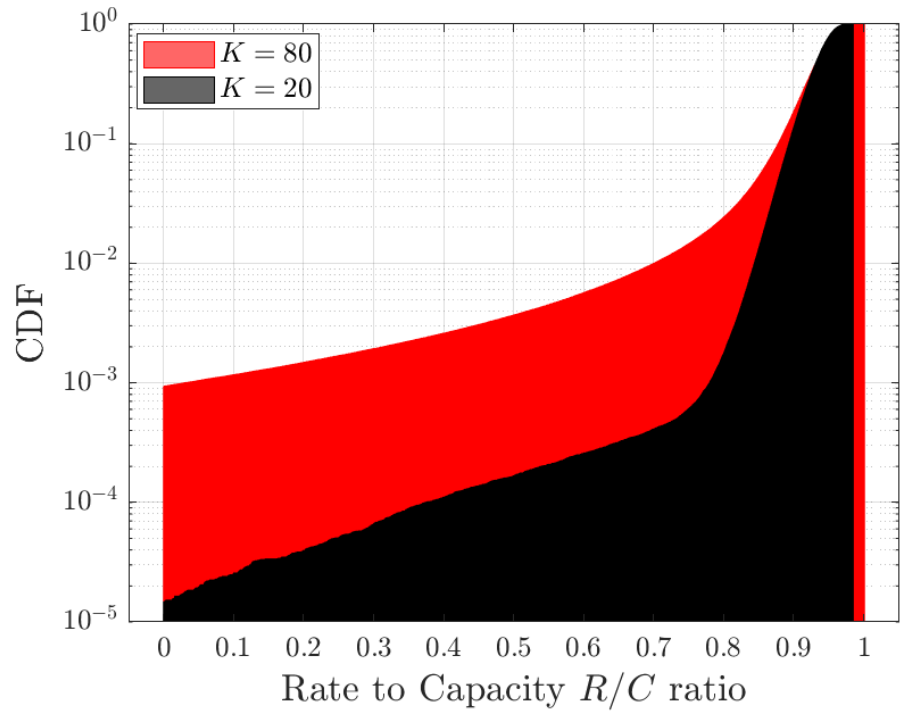}
\par\end{centering}
\caption{Histogram of the experienced rate to capacity ratio, for various densities
of VUEs $K$ with deterministic arrivals, finite blocklength case.\textcolor{blue}{\label{fig:RateCapacityRatio}}}
\end{figure}

In Fig.\,\ref{fig:RateCapacityRatio}, we validate the accuracy of
using normal approximation within our proposed framework. According
to \cite[Section IV-C]{Polyanskiy2010}, the normal approximation
is quite accurate when transmitting at a large fraction of the channel
capacity (e.g., 0.8 of channel capacity). Fig.\,\ref{fig:RateCapacityRatio}
shows the CDF histogram of the rate to capacity ratio experienced
within our simulations. Note that the transmission rate is equal to
or larger than $0.8$ of the channel capacity about $99\%$ of the
time, which verifies the accuracy of using normal approximation within
our proposed framework.

\subsection{Performance Comparison Based on AoI}

\begin{figure}
\begin{centering}
\centering\includegraphics[width=0.8\columnwidth]{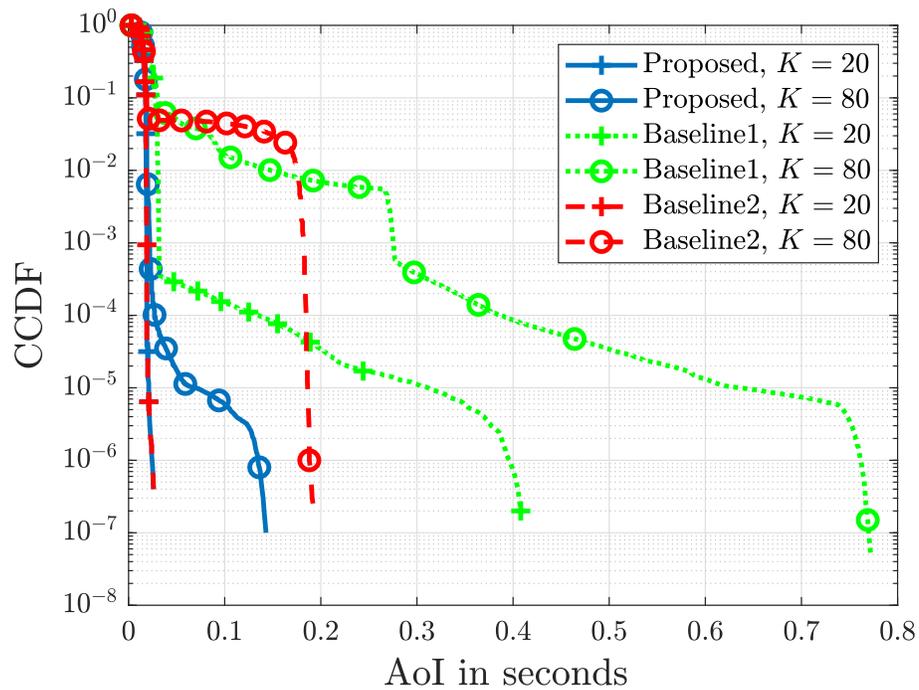}
\par\end{centering}
\caption{CCDF of the AoI for various densities of VUEs $K$ with deterministic
arrivals, for large blocklength.\label{fig:CCDF_AoIDG1}}
\end{figure}
\begin{figure}
\begin{centering}
\centering\includegraphics[width=0.8\columnwidth]{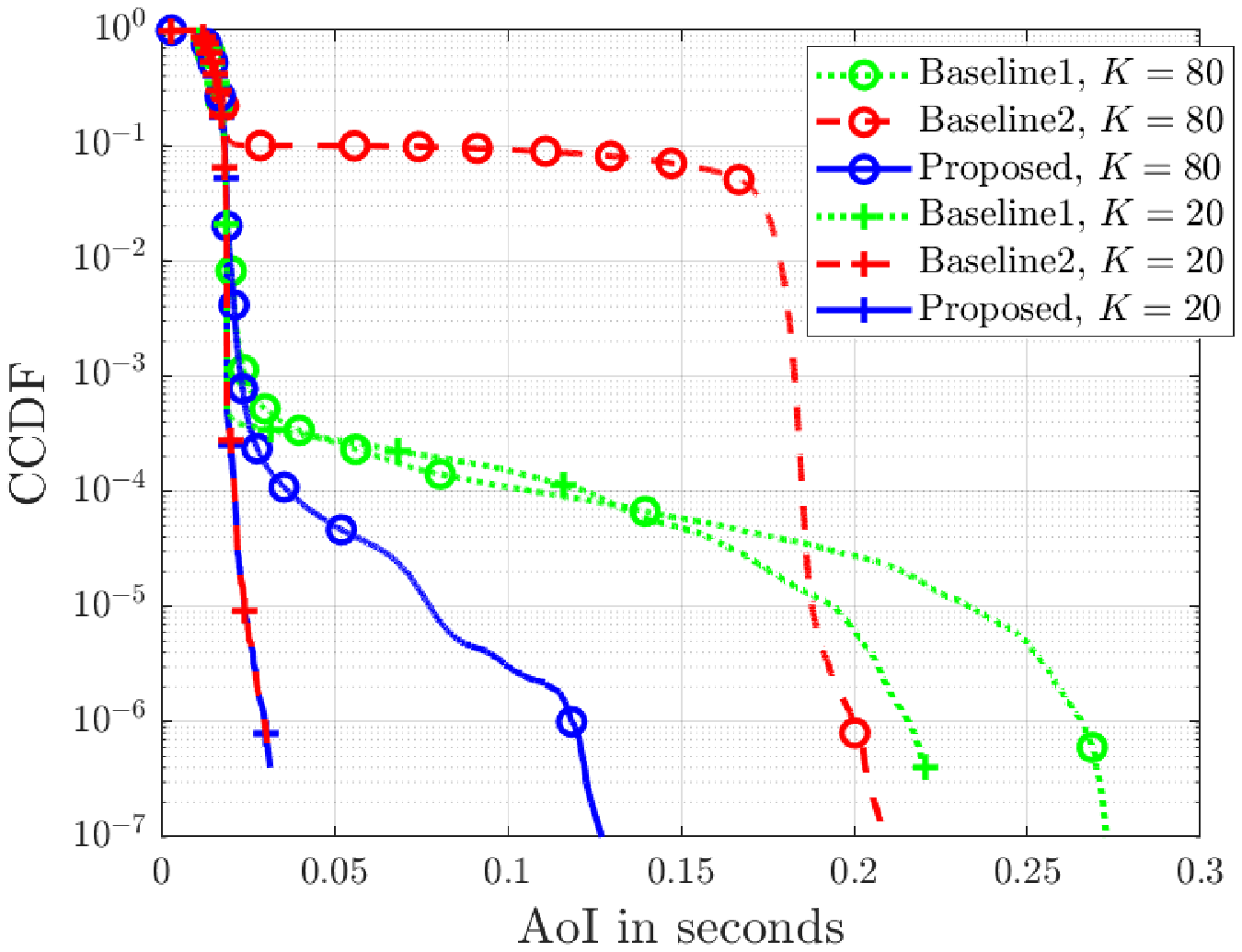} 
\par\end{centering}
\caption{CCDF of the AoI for various densities of VUEs $K$ with deterministic
arrivals, for finite blocklength with $L=550$.\textcolor{blue}{\label{fig:CCDF_AoIDG1Short}}}
\end{figure}
In Fig.\,\ref{fig:CCDF_AoIDG1} and \ref{fig:CCDF_AoIDG1Short},
the CCDFs of the AoI are plotted for the deterministic arrivals with
large and finite blocklengths, with a comparison to \emph{Baseline
1} and \emph{Baseline 2}. Note that, by using \eqref{eq:AoICalculation}
the AoI can be calculated at the beginning of every time slot and
then used to plot the CCDF. In particular, Fig.\,\ref{fig:CCDF_AoIDG1}
and \ref{fig:CCDF_AoIDG1Short} show the AoI distributions for two
densities of VUEs with large blocklength and finite blocklength ($L=550$
channel uses (CU)), respectively. From these two figures, we can see
that the AoI performance for the proposed method outperforms the baseline
models for both VUE densities, yielding improved reliability, except
for \emph{Baseline 2} when $K=20$. For $K=20$, due to the low VUE
density, each transmitter-receiver pair maintains high data rate yielding
low-to-no events where AoI exceeding the threshold. Therefore, the
constraints modeled using EVT to control the extreme events have no
impact in which the proposed and \emph{Baseline 2} exhibit similar
AoI distributions. Note that, as VUE density increases to $K=80$,
increased interference and lower rates results in events with AoI
exceeding the threshold. Therein, the proposed EVT-based constraints
actively contribute to maintain the extreme AoI events whereas \emph{Baseline
2}\textbf{ }has no control on such extreme events. As a result, the
proposed method yields reduced AoI over the network compared to \emph{Baseline
2}\textbf{.} From Fig.\,\ref{fig:CCDF_AoIDG1}, we observe that,
when $K=80$, \emph{Baseline 1} experiences at least $0.6\,s$ in
AoI with $10^{-5}$ probability, while the proposed model experiences
less than $0.1\,s$ for the same probability. Also, the probability
of having the AoI greater than $0.1\text{ s}$ is more than $10^{-2}$
in the case of \emph{Baseline 2}, while it is less than $10^{-5}$
in the proposed approach. In short, our proposed approach succeed
in managing the AoI tail compared to the baselines.

\subsection{Impact of the Lyapunov Tradeoff Parameter $V$}

\begin{figure}
\begin{centering}
\centering\includegraphics[width=0.8\columnwidth]{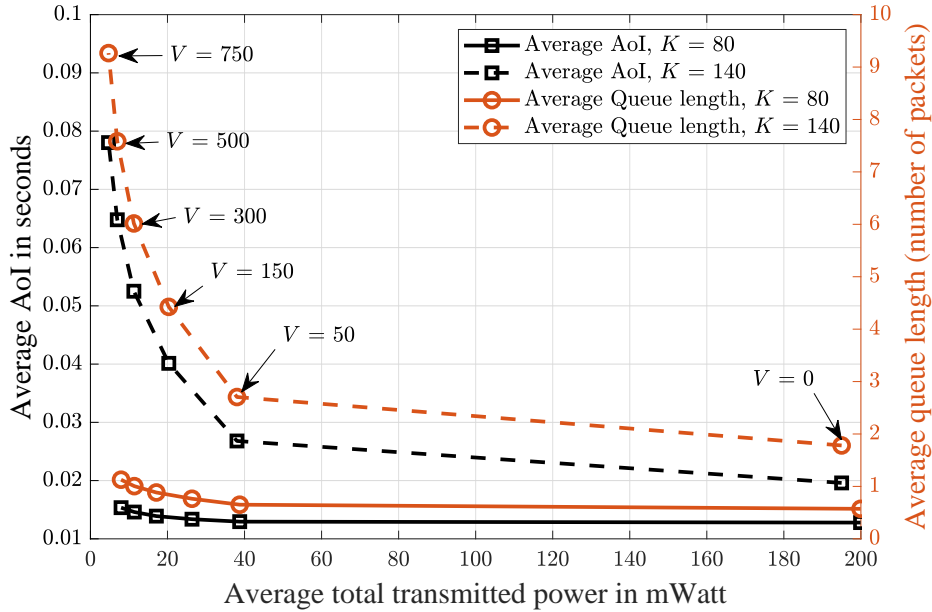} 
\par\end{centering}
\caption{Transmit power, average AoI, and queue length trade-off for various
VUEs densities $K$ with Markovian arrivals for large blocklength.\label{fig:Lyapunov_Tradeoff}}
\end{figure}
\begin{figure}
\begin{centering}
\centering\includegraphics[width=0.8\columnwidth]{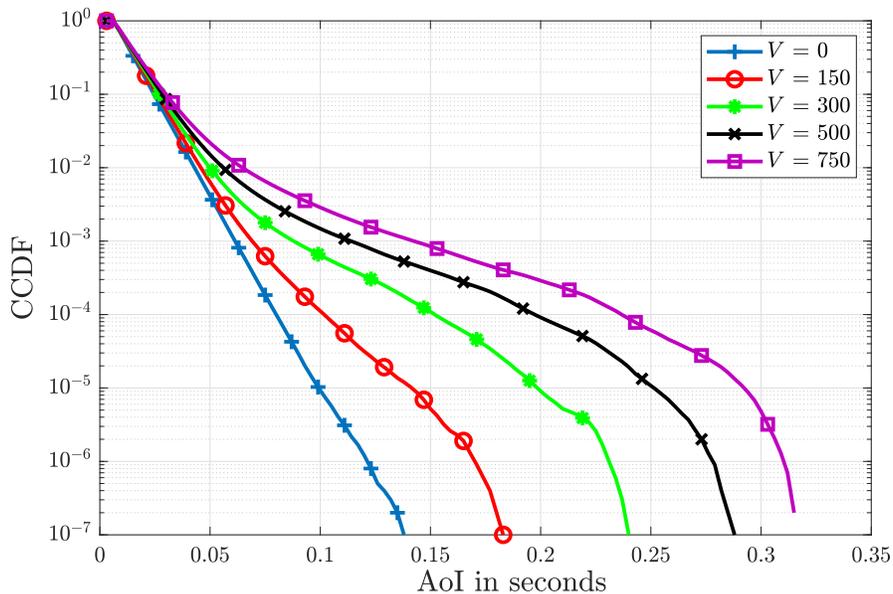}
\par\end{centering}
\caption{The AoI's CCDF for various Lyapunov parameter $V$ when $K=80$ with
Markovian arrivals for large blocklength.\label{fig:CCDF_LyapunovAoI}}
\end{figure}
Next, we discuss the impact of Lyapunov control parameter $V$ on
the AoI and the optimal transmit power. In Luapunov optimization,
$V$ controls the tradeoff between transmit power and queue stability.
Therefore, the impacts of $V$ on AoI, queue length, and the transmission
power on average are analyzed in Fig.\,\ref{fig:Lyapunov_Tradeoff}
for two different VUE densities $K=\{80,140\}$ with the Markovian
arrival case and large blocklength. It can be noted that, when $V$
is small, the priority of the VUEs is to minimize the physical and
virtual queue lengths (i.e. maximize data rates) rather than minimizing
the power consumption. Therefore, for small $V$, a smaller AoI can
be observed at the price of an increased power consumption as illustrated
in Fig.\,\ref{fig:Lyapunov_Tradeoff}. In contrast, a large $V$
ensures a reduced power consumption with increased AoI. Fig.\,\ref{fig:Lyapunov_Tradeoff}
shows that the average AoI and queue length of all VUE pairs in the
Markovian arrival case increases as the Lyapunov parameter $V$ increases,
while the total average transmitted power decreases.

Fig.\,\ref{fig:CCDF_LyapunovAoI} shows the CCDFs of AoI in the Markovian
arrival case for different $V$. Similar to the average AoI, Fig.\,\ref{fig:CCDF_LyapunovAoI}
validates our claim on the performance loss of overall AoI with increasing
$V$. Although similar behavior can be observed for the deterministic
arrival case, the differences are insignificant, in which the results
are not presented.

\subsection{Impact of the Arrival Rate}

\begin{figure}
\begin{centering}
\centering\includegraphics[width=0.8\columnwidth]{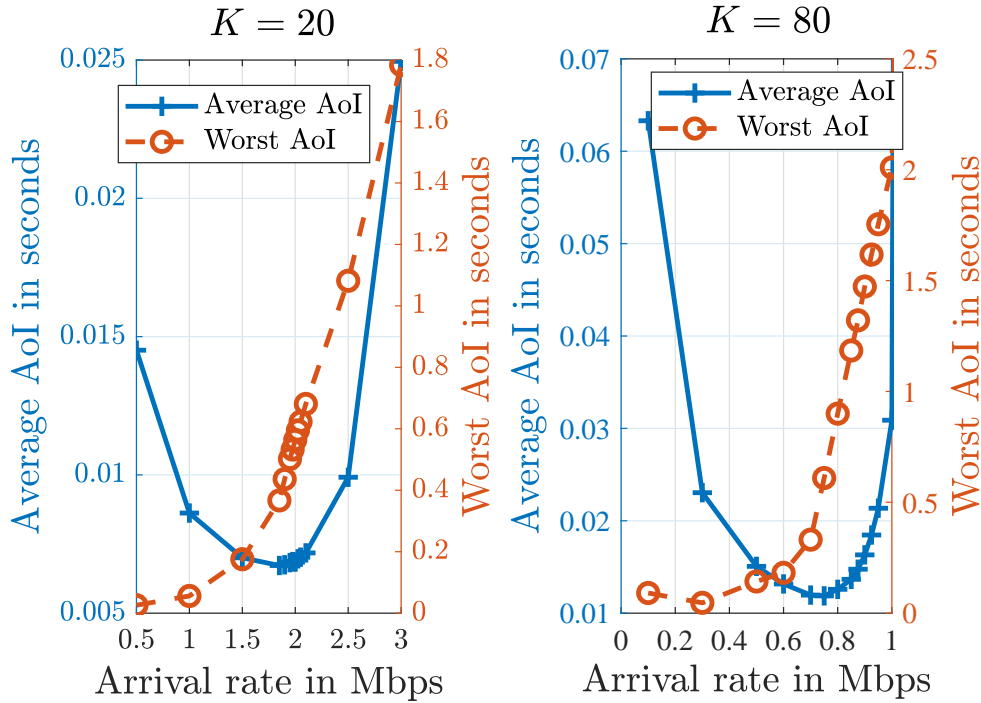} 
\par\end{centering}
\caption{Arrival rate versus AoI trade-off, with $K=80$ and $K=20$ VUEs in
the deterministic arrival case with large blocklength.\label{fig:tradeoffArrivalrate}}
\end{figure}
This subsection studies the impact of the status updates' arrival
rate for two different VUE densities, $K=80$ and $K=20$, for both
average and worst AoI. The \emph{worst AoI} is defined as the maximum
AoI experienced by all VUEs during the simulation duration. Fig.\,\ref{fig:tradeoffArrivalrate}
shows that, when the arrival rate is less than $0.1\text{ Mbps}$
($0.5\text{ Mbps}$) when $K=80$ ($20$), the network experiences
a higher average AoI. It also shows that increasing the arrival rate
will lead to a better average AoI, up to a certain point (around $0.8\text{ Mbps}$
when $K=80$ and $2\text{ Mbps}$ when $K=20$), after which the average
AoI starts increasing again. This happens because, at low arrival
rates, queue stability is easily achieved. Thus, there is no rush
of emptying the queue at the transmitter and data packets are more
probable to be queued for a longer period, leading to a higher average
AoI. When arrival rate increases, data packets within the queue are
transmitted more frequently to maintain the queue stability. Thus,
achieving lower average AoI. Further increasing the arrival rate will
make it difficult to stabilize the queues. Hence, packets are lingered
in the queue for longer duration, leading to the increased average
AoI. Note that, at low VUE density ($K=20$), the network can withstand
higher arrival rates. Fig.\,\ref{fig:tradeoffArrivalrate} also highlights
that the worst AoI increases slightly with the arrival rate up to
the aforementioned optimal arrival rate for the average AoI, and exhibits
a rapid increase afterwards. Moreover, we note that the worst AoI
is about 10 fold higher than the average AoI. Although the AoI achieves
a small average, the worst AoI is heavy-tailed. In this situation,
relying on the average AoI is inadequate for ensuring URLLC. This
discrepancy between these two metrics demonstrates that the tail characterization
is instrumental in designing and optimizing URLLC-enabled V2V networks.

\subsection{Impact of Blocklength $L$ and Block Error Probability $\varepsilon$}

\begin{figure}
\begin{centering}
\centering\includegraphics[width=0.8\columnwidth]{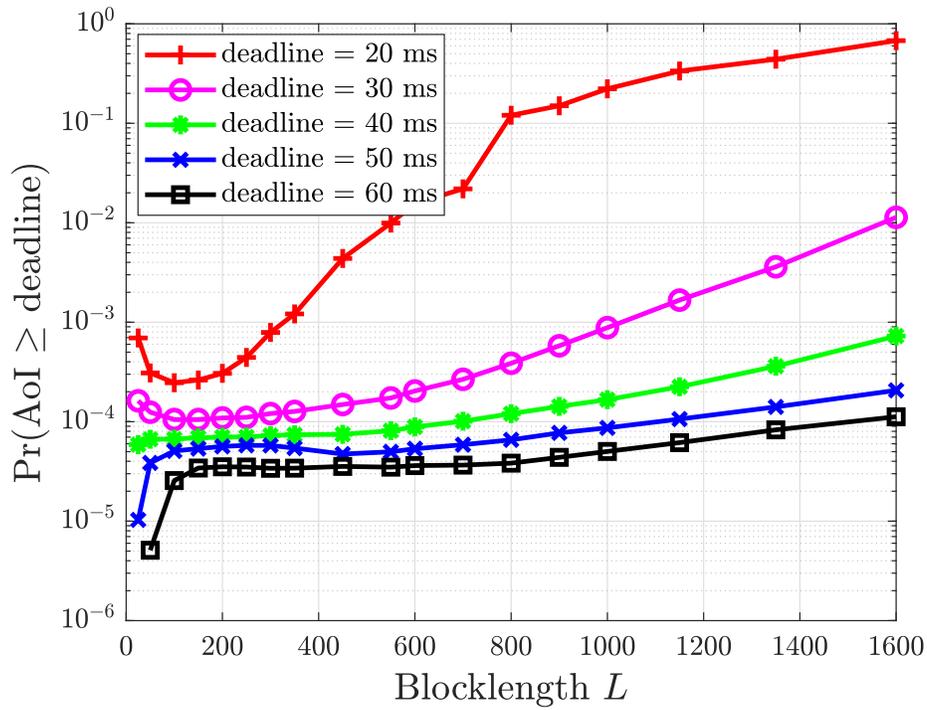} 
\par\end{centering}
\caption{Blocklength $L$ vs. AoI violation probability, $K=80$ VUEs in the
deterministic arrival case.\label{fig:blocklength}}
\end{figure}
\begin{figure}
\begin{centering}
\centering\includegraphics[width=0.8\columnwidth]{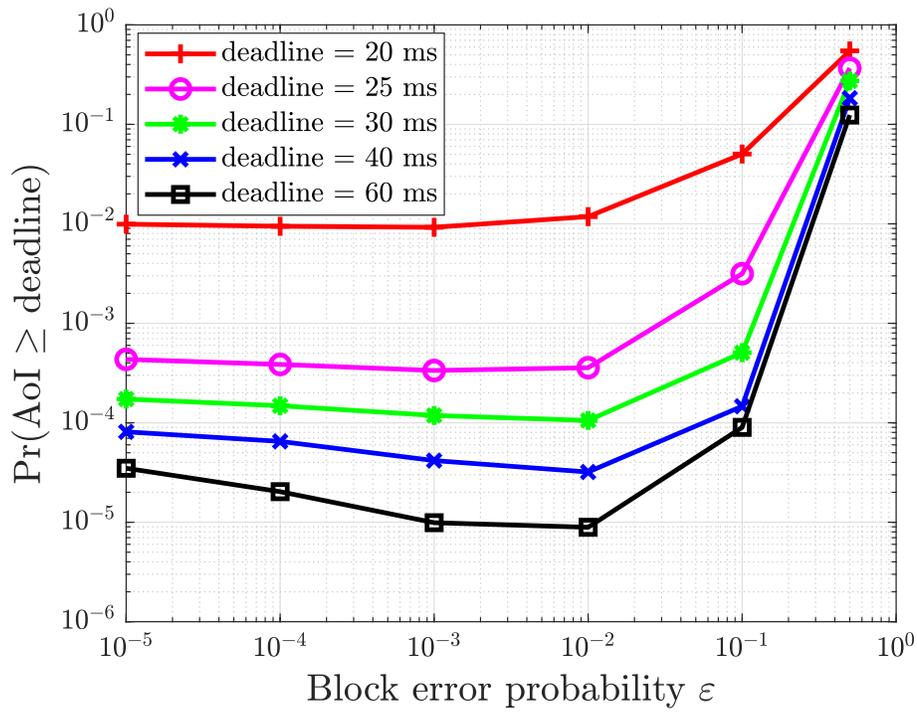}
\par\end{centering}
\caption{AoI violation probability vs. block error probability $\varepsilon$,
$K=80$ VUEs in the deterministic arrival case.\label{fig:blockerror}}
\end{figure}
Fig.\,\ref{fig:blocklength} illustrates the effect of changing the
blocklength $L$ on the probability of AoI exceeding a set of deadlines,
$\textrm{Pr}\left\{ \text{AoI}\geq\text{deadline}\right\} $. For
a given $L$, e.g., when $L=550$, $\textrm{Pr}\left\{ \text{AoI}\geq\text{deadline}\right\} $
can be computed from Fig.\,\ref{fig:CCDF_AoIDG1Short}, where $\textrm{Pr}\left\{ \text{AoI}\geq\text{deadline}\right\} $
is the actual achieved probability. Since the blocklength $L$ is
determined by the bandwidth $\omega$ and the time slot duration (coherence
time) $\tau$ as per $L=\omega\tau$, $L$ is varied by changing the
coherence time $\tau$ (i.e., vehicle speed \cite{CoherenceTimeandVelocity}).
Fig.\,\ref{fig:blocklength} shows that there exists a blocklength
$L$ above which the AoI violation probability increases with increasing
the blocklength $L$. Within this region, when the blocklength $L$
increases, both the transmission rate and transmit power consumption
increase yielding more interference and, hence, the increase in the
AoI violation probability. However, within the region where the blocklength
and deadline are small, the small blocklength $L$ results in a small
transmission rate and, hence, the packets are accumulated within the
queue yielding a higher AoI violation probabilities. Moreover, for
a fixed blocklength $L$, the AoI violation probability decreases
with increasing the deadline. Finally, we show the impact of changing
the block error probability $\varepsilon$ on the AoI violation probability
in Fig.\,\ref{fig:blockerror}. Note that lowering the block error
probability $\varepsilon$ decreases the transmission rate $R$ but
increases the AoI. When $\varepsilon$ is large, e.g., $\varepsilon\in[10^{-2},0.5]$,
the AoI violation probability is mainly caused by unsuccessful packet
decoding. Thus, lowering $\varepsilon$ decrease the AoI violation
probability. When $\varepsilon$ is below $10^{-3}$, the effect of
unsuccessful packet decoding is not significant. In this regime, AoI
violation is caused by the low transmission rate which results in
a high AoI. Due to this reason, lowering $\varepsilon$ increases
the AoI violation probability.

\section{Conclusion and future work\label{sec:Conclusion}}

In this paper, we have studied the problem of ultra-reliable and low-latency
vehicular communication, considering both deterministic and stochastic
arrivals . For this purpose, we have first defined a new reliability
measure, in terms of probabilistic AoI, and have established a novel
relationship between the AoI and queue-related probability distributions.
Then, we have shown that characterizing the AoI tail distribution
can be effectively done using EVT. Subsequently, we have formulated
a transmit power minimization problem subject to the probabilistic
AoI constraints and solved it using Lyapunov optimization. Furthermore,
we have studied the impact of short packets and how it affects the
optimization of AoI. Simulation results have shown that the proposed
approach yields significant improvements in terms of AoI and queue
length, when compared to baseline models. Moreover, an interesting
tradeoff between the status updates' arrival rate and the average
and worst AoI achieved by the network has been exposed. We have also
shown the existence of a blocklength at which the AoI violation probability
is minimized. Many future extensions can be considered as follows:
packet retransmissions, dynamic vehicle association, vehicles' handover
between RSUs, considering a multicast/broadcast scheme and finally,
incorporating different queuing models and policies.

\appendices{}

\section{Proof of Lemma \ref{lem:Assuming-that-packet }\label{sec:AppendixA}}

Since $T=\tau(t+1)$, then $\mathrm{Pr}\left\{ T_{k}^{\text{D}}(\hat{\imath})>\tau(t+1)\right\} =\mathrm{Pr}\left\{ \hat{\imath}\text{ is NOT served before time }\tau(t+1)\right\} $,
which means that $\hat{\imath}$ is not served at or before time slot
$t$. Subsequently, we apply \eqref{eq:served_indices} and derive
\begin{align*}
 & \mathrm{Pr}\left\{ T_{k}^{\text{D}}(\hat{\imath})>\tau(t+1)\right\} \\
 & \overset{(a)}{=}\mathrm{Pr}\left\{ \hat{\imath}>tA-1-\max\left(Q_{k}(t)-R_{k}(t),0\right)\right\} \\
 & \leq\mathrm{Pr}\left\{ \frac{A}{\tau}(\tau(t+1)-d_{\text{D}})+1>tA-1-\left(Q_{k}(t)-R_{k}(t)\right)\right\} \\
 & =\mathrm{Pr}\left\{ Q_{k}(t)>R_{k}(t)-\psi\right\} ,
\end{align*}
where $\hat{\imath}=\lceil\frac{A}{\tau}(T-d_{\text{D}})\rceil\leq\frac{A}{\tau}(\tau(t+1)-d_{\text{D}})+1$
is used in step \emph{(a)}. It also should be noted that if $\hat{\imath}$
departs after $\tau(t+1)$, then $Q_{k}(t)-R_{k}(t)>0$, which is
used in the same step.

\section{Proof of Lemma \ref{lem:Given-an-M/G/1}\label{sec:AppendixB}}

Let $\check{\imath}$ be the last packet that departed at or just
before time $T$. Thus, $\Delta_{k}(T)=T-T_{k}^{\text{A}}(\check{\imath})$.

\textbf{\emph{Case 1:}} If no arrivals occurred during the time interval
$\left[T-d_{\text{M}},T\right)$, then the arrival time of $\check{\imath}$
must be strictly less than $T-d_{\text{M}}$, i.e., $T_{k}^{\text{A}}\left(\check{\imath}\right)<T-d_{\text{M}}$.
Therefore, 
\[
\mathrm{Pr}\left\{ \Delta_{k}(T)>d_{\text{M}}\right\} =\mathrm{Pr}\left\{ T-T_{k}^{\text{A}}\left(\check{\imath}\right)>d_{\text{M}}\right\} =1.
\]

\textbf{\emph{Case 2:}} If at least one arrival occurred during the
time interval $\left[T-d_{\text{M}},T\right)$, then $T_{k}^{\text{A}}\left(\hat{\imath}\right)$
is bounded between $T-d_{\text{M}}\leq T_{k}^{\text{A}}\left(\hat{\imath}\right)<T$.
Since $\hat{\imath}$ is the first arrival on or after time $T-d_{\text{M}}$,
in this case, we need to show that the event $\left\{ \Delta_{k}(T)<d_{\text{M}}\right\} $
is equivalent to the event $\left\{ T_{k}^{\text{D}}(\hat{\imath})<T\right\} $
as in \cite{champatiAoI}. If the event $\left\{ \Delta_{k}(T)<d_{\text{M}}\right\} $
occurred, then $T_{k}^{\text{A}}\left(\check{\imath}\right)\geq T-d_{\text{M}}$.
By definition of $\hat{\imath}$, we should have $T_{k}^{\text{A}}\left(\hat{\imath}\right)\leq T_{k}^{\text{A}}\left(\check{\imath}\right)$
which entails that $T_{k}^{\text{D}}\left(\hat{\imath}\right)\leq T_{k}^{\text{D}}\left(\check{\imath}\right)\leq T$,
due to FCFS assumption. As a result, $\left\{ \Delta_{k}(T)<d_{\text{M}}\right\} \subseteq\left\{ T_{k}^{\text{D}}(\hat{\imath})<T\right\} $.

To show equivalence of the two events, we show that the previous relation
also holds the other way round. If event $\left\{ T_{k}^{\text{D}}(\hat{\imath})<T\right\} $
occurred, we must have $T_{k}^{\text{A}}\left(\hat{\imath}\right)\leq T_{k}^{\text{A}}\left(\check{\imath}\right)$.
Otherwise, $T_{k}^{\text{D}}(\check{\imath})<T_{k}^{\text{D}}(\hat{\imath})\leq T$
which contradicts the definition of $\check{\imath}$. Therefore,
$\Delta_{k}(T)=T-T_{k}^{\text{A}}\left(\check{\imath}\right)\leq T-T_{k}^{\text{A}}\left(\hat{\imath}\right)\leq T-(T-d_{\text{M}})=d_{\text{M}}$.
This implies that $\left\{ T_{k}^{\text{D}}(\hat{\imath})<T\right\} \subseteq\left\{ \Delta_{k}(T)<d_{\text{M}}\right\} $
and therefore, $\left\{ \Delta_{k}(T)<d_{\text{M}}\right\} \equiv\left\{ T_{k}^{\text{D}}(\hat{\imath})<T\right\} $.

Note that, due to the Poisson arrivals assumption. Using (\ref{eq:Poisson}),
the probability of no arrivals occurring during the time interval
$\left[T-d_{\text{M}},T\right)$ (Case 1) is $e^{-\frac{\lambda d_{\text{M}}}{\tau}}$,
while the probability of at least one arrival during the same interval
(Case 2) is $1-e^{-\frac{\lambda d_{\text{M}}}{\tau}}$. Therefore,
$\mathrm{Pr}\left\{ \Delta_{k}(T)>d_{\text{M}}\right\} =1\cdot e^{-\frac{\lambda d_{\text{M}}}{\tau}}+\mathrm{Pr}\left\{ T_{k}^{\text{D}}(\hat{\imath})>T\right\} \left(1-e^{-\frac{\lambda d_{\text{M}}}{\tau}}\right)$.

\section{Proof of Lemma \ref{lem:MapMG1} \label{sec:AppendixC}}

Since $\hat{\imath}$ is the packet that first arrives on or after
time $T-d_{\text{M}}$, then $\mathrm{Pr}\left\{ T_{k}^{\text{D}}(\hat{\imath})>T\right\} =\mathrm{Pr}\left\{ \text{Sojourn time}>d_{\text{M}}\right\} $,
where the sojourn time is the total time a packet spend within the
system.

Observing the time slot $t$ and given that $R_{k}(t)$ is the number
of packets that will be served during this slot, i.e. during period
$\left[\tau t,\tau(t+1)\right]$. Therefore, $\left\{ \text{Sojourn time}>d_{\text{M}}\right\} $
if the number of packets arrived during time period $\left[\tau t-d_{\text{M}},\tau(t+1)-d_{\text{M}}\right]$
is more than $R_{k}(t)$.

From \eqref{eq:Poisson}, the Poisson distribution is irrelevant to
the start and end times of a given period, therefore, $\mathrm{Pr}\biggl\{\text{Number of packets arrived during time period }\Bigl[\tau t-d_{\text{M}},\tau(t+1)-d_{\text{M}}\Bigr]>R_{k}(t)\biggr\}=\mathrm{Pr}\left\{ A_{k}(t)>R_{k}(t)\right\} =\mathrm{Pr}\left\{ \text{Sojourn time}>d_{\text{M}}\right\} $.

\bibliographystyle{IEEEtran}
\bibliography{References,Response_nocolor}

\end{document}